\documentclass[english,aps,pra,superscriptaddress,twocolumn,floatfix]{revtex4}
\usepackage[T1]{fontenc}
\usepackage[latin9]{inputenc}
\usepackage{units}
\usepackage{amssymb}
\usepackage{graphicx}
\usepackage{color}
\usepackage{natbib}
\usepackage[caption=false]{subfig}
\usepackage{babel}
\usepackage{array}
\usepackage{multirow}
\usepackage{placeins}
\usepackage{hyperref}
\graphicspath{./Figures/}

\newcommand{\ket}[1]{
  \vert #1  \rangle
}
\newcommand{\bra}[1]{
  \langle #1 \vert
}

\begin{document}

\title{Freely Scalable Quantum Technologies\\ using Cells of 5-to-50 Qubits\\ with Very Lossy and Noisy Photonic Links}

\author{Naomi H. Nickerson}
\affiliation{Department of Physics, Imperial College London, Prince Consort Road, London SW7 2AZ, UK } 

\author{Joseph F. Fitzsimons}
\affiliation{Singapore University of Technology and Design, 20 Dover Drive, Singapore 138682}
\affiliation{Center for Quantum Technologies, National University of Singapore, Block S15, 3 Science Drive 2, Singapore 117543}
	
\author{Simon C. Benjamin}
\affiliation{Department of Materials, University of Oxford, Parks Road, Oxford OX1 3PH, UK}

\begin{abstract}
Exquisite quantum control has now been achieved in small ion traps, in nitrogen-vacancy centres and in superconducting qubit clusters. We can regard such a system as a {\em universal cell} with diverse technological uses from communication to large-scale computing, provided that the cell is able to network with others and overcome any noise in the interlinks. Here we show that loss-tolerant entanglement purification makes quantum computing feasible with the noisy and lossy links that are realistic today: 
 With a modestly complex cell design, and using a surface code protocol with a network noise threshold of $13.3\%$, we find that interlinks which attempt entanglement at a rate of $2$~MHz but suffer $98\%$ photon loss can result in kilohertz computer clock speeds (i.e. rate of {\em high fidelity} stabilizer measurements). Improved links would dramatically increase the clock speed. 
Our simulations employed local gates of a fidelity already achieved in ion trap devices. \\
\end{abstract}
\maketitle

Within the last year there have been remarkable advances in the fidelity with which small quantum devices can be controlled. The two most mature systems are ion traps and superconducting qubits. In ion trap devices single qubit fidelities~\cite{LucasPreP} have reached $99.9999\%$, with combined preparation and measurement of $99.93\%$. Moreover two-qubit operations~\cite{IonBest2Q} have been reported with fidelities up to $99.9\%$. Meanwhile a superconducting qubit device (SQD) containing five qubits~\cite{MartinisNew} has been demonstrated with all qubit manipulations above $99.3\%$. At the same time there has been rapid progress in the study of nitrogen vacancy centres in diamond -- single electron spin manipulation is possible with $99\%$ fidelity~\cite{Wrachtrup14}, and it is possible to manipulate nuclei that are relatively far from the centre, so that each NV centre may be thought of as a group of several qubits interacting with an optically active core~\cite{distantNuclei}.

\begin{figure}[b]
\centering{}
\includegraphics[width=0.98\columnwidth]{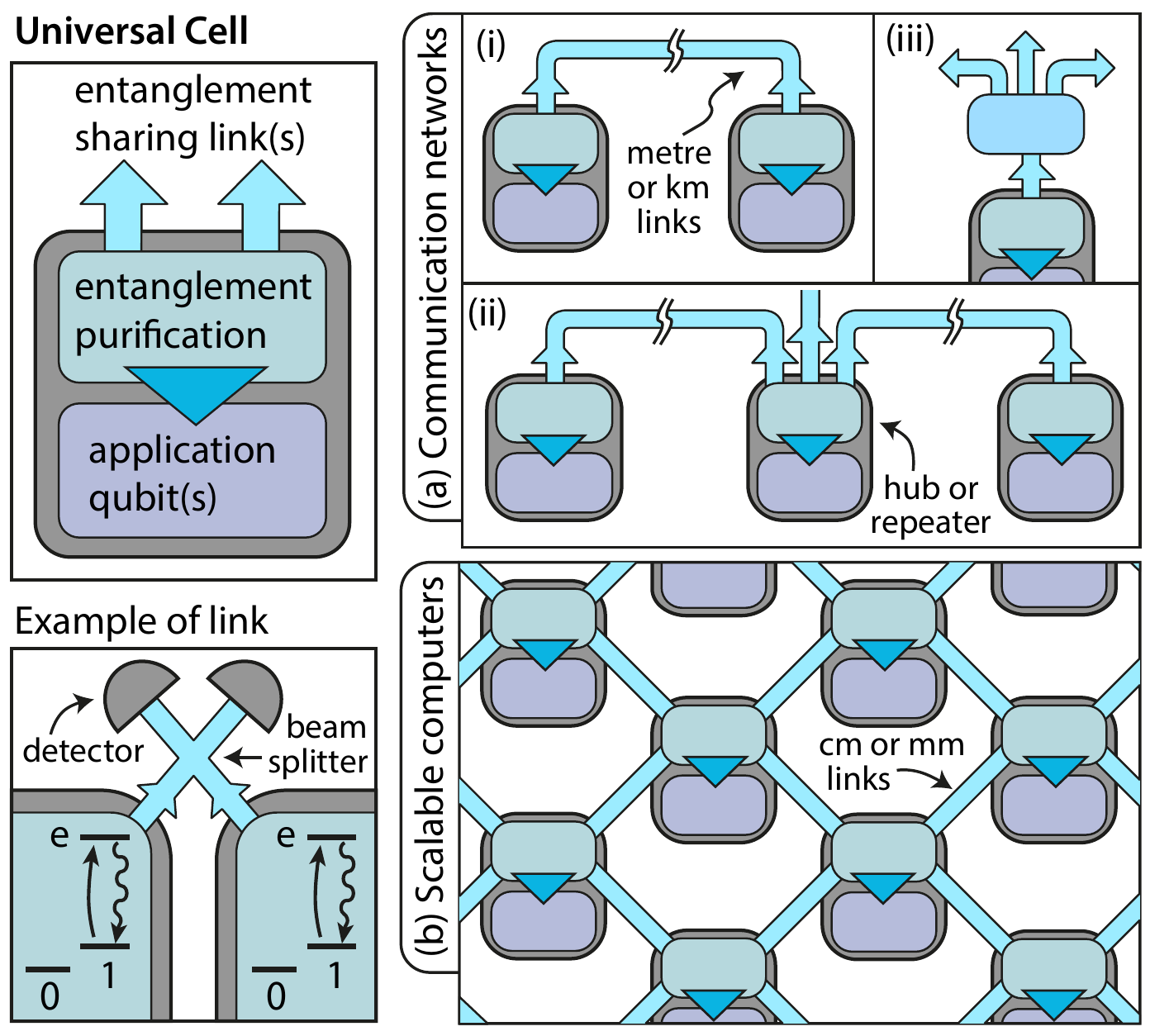}
\caption{\label{fig:paradigm}
A small, well-controlled  quantum system interfaced to a noisy entanglement-sharing channel constitutes a {\bf universal cell} if it can purify the  entanglement to a high fidelity. Such cells enable secure communication; monogamy of entanglement~\cite{monogEnt} means the {\em links need not be secure} and may be either direct (i), or via repeater hubs (ii) or switches (iii). (b) Moreover a dense array of cells bridged by short links constitutes a freely-scalable computer, as we analyse here.
}
\end{figure}

These prototype systems are small; none of them contain as many as $20$ qubits. But importantly in each case it is possible to bridge between small systems using photonic channels, albeit with lower entanglement fidelities and in a probabilistic way that may require many attempts. In the ion trap community there are well established methods for entangling ions in separate traps, and recent progress associated with projects such as the MUSIQC initiative~\cite{MUSIQC} have led to successful entanglement at a rate of hertz~\cite{monroeNew}. This can be improved by orders of magnitude by hardware advances and by loss-adapted protocols, as we describe presently in this paper. In SQDs a well established means of interfacing qubits is to exploit microwave photons in cavities~\cite{SQDcavityCouple}. This suffices for short range bridging and moreover remote entanglement of two superconducting qubits separated by more than a meter of coaxial cable has recently been demonstrated~\cite{Roch_RemoteEntanglement}. In the case of NV centre research, successful optical linking of qubits can occur either within the same sample~\cite{Wrachtrup14} or over metres of separation~\cite{BERNIEN_2012nvEntanglement}, and teleportation~\cite{DelftTeleport} with fidelity around $86\%$ as been achieved.

Thus the quantum state of the art includes well-controlled small groups of qubits, which we refer to as `cells' in this paper, together with inter-cell entanglement links that may be non-deterministic and, even when successful, noisy. These ingredients may already suffice to develop fully scalable technologies: although the fidelities over the links are too low to directly enable secure communication or fault tolerant computing, crucially the fidelities {\em within} cells are now high enough to support entanglement purification. This process allows one to improve the fidelity of a quantum channel by combining several successive uses of the link. Thus at the cost of lowering the effective bandwidth we have a powerful paradigm in which small cells link to one another through a kind of internal digital filter where purification is performed, see Fig.~\ref{fig:paradigm}. This paradigm universally supports quantum technologies on any scale. On the large scale, when the bridges between cells are metres or kilometres long, cellular nodes enable secure communication and other distributed information tasks. However the present paper concerns freely scalable quantum computing, where the optical bridges connect a dense array of cells with spacings on the order of centimetres or less.

Any technology based on high performing cells bridged by very imperfect links will only be practical if entanglement purification is efficient and robust. The protocol should have frugal requirements for `work space' qubits within a cell, it should require only achievable levels of fidelity for local gates and measurements, and most importantly it should minimise the time cost by requiring only a few uses of the noisy quantum channel (i.e. a small number of low fidelity Bell pairs) in order to purify a high fidelity shared state. These desiderata are in tension with one another, and of course the achievable values will depend on the fidelity of the native channel as well as the target fidelity that enables the task in question (e.g. communication or computation). Furthermore certain tasks, such as the stabilizer based computing considered here, are best enabled by multiparty entangled states that are more complex than simple Bell pairs. 

A seminal paper in the purification literature is that of Briegel and D\"ur~\cite{BRIEGEL_entanglementPurification} which showed that by using a tiering system, one can promote even very noisy `raw' entanglement to a fidelity that is {\em of the order of the fidelity of the local operations}. A number of authors have extended this idea, for example through a means to purify phase noise very efficiently~\cite{EarlPRA} and by introducing the idea of `double selection' Ref.~\cite{FUJII_doubleSelection}, which was then used in a scheme~\cite{FUJII_distributedArchitechture} that can tolerate channel noise up to $30\%$ in the context of quantum computing. Recent work has even pushed the acceptable limits of local noise to comparable levels~\cite{BriegelNew}. Here we employ a range of such techniques in order to obtain what we believe to be the most practical purification protocols yet described for the context of quantum computing. Moreover we show how these protocols can guide specific hardware design to achieve optimum efficiency. We aim to determine whether the intra-cell operations and limited inter-cell links that are regarded as {\em achievable today} can suffice for full scale quantum computing, assuming that the various accomplishments that have been made in different (but compatible) experiments can be engineered into a single platform. We conclude that the answer is yes.

\begin{figure*}
\centering{}
\includegraphics[width=2.06\columnwidth]{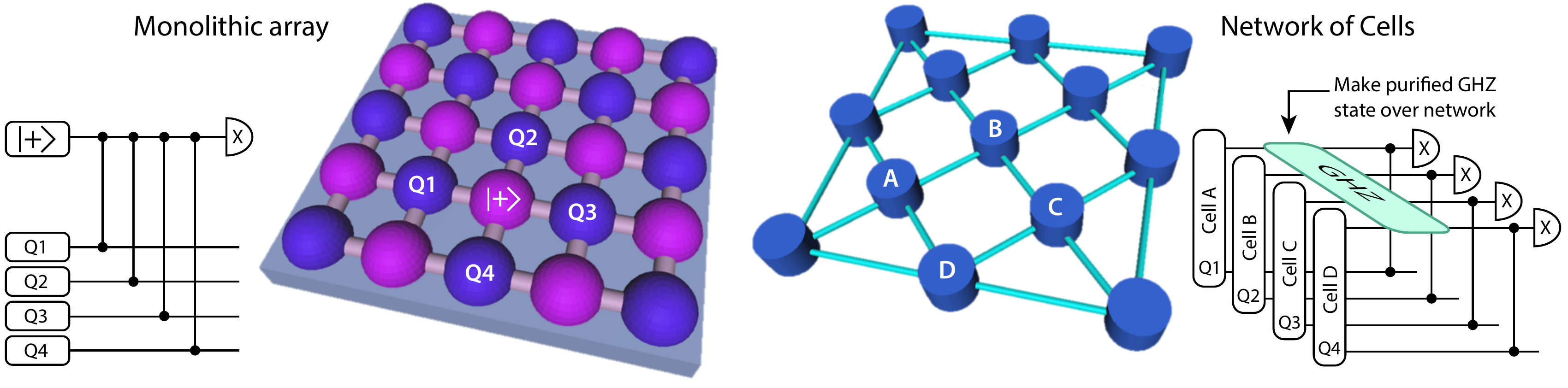}
\caption{\label{fig:Monolithic vs network}
\textbf{Monolithic vs. Network quantum computing} 
A piece of surface code of dimension $L=2$ is
represented in the monolithic picture (left) and in a network
architecture (right). In a monolithic structure~\cite{FOWLER_2009high,WANG_QCwithNNinteractionsAndErrorRatesOverOnePercent} all qubits are contained
within one physical system and two-qubit gates are performed directly
between qubits via some physical interaction. In the network picture
the system is instead divided up into small cells, each of which
contains a modest number of qubits which interact directly, while between cells 
only noisy and lossy interactions are possible. For the monolithic system, stabilizer 
measurements are performed by using a dedicated set of ancilla qubits (pink) 
interlaced with the data qubits (blue). In contrast, an efficient route to making 
stabilizer measurements in the network model is to first purify a shared GHZ 
state between the cells involved, and then use this resource to evaluate the stabilizer: 
The parity of the four qubits measured out from the GHZ tells
us the stabilizer outcome. In subsequent cycles, cells are grouped
into different sets of four in order to evaluate a complementary set
of stabilizers, see Appendix~\ref{sub:scheduling}. }
\end{figure*}

Our approach here is an evolution of the scheme in Ref.~\cite{nickerson_topological} which was designed to fight network \textit{noise}. We will extend that scheme to be efficient versus severe \textit{loss} while retaining the noise tolerance, and this allows us to analyse the `clock speed' of the resulting computer. Our approach requires a total of at  least five qubits per cell: four qubits for purification of noise on the cell-cell coupling links, and one that is involved in the actual quantum computation (a so-called `data qubit' as explained presently). Ion traps, SQDs and NV centre systems can all scale to five qubits. However with ion traps and SQDs we may eventually have the luxury of tens of qubits per cell. In that case we can make good use of the additional structure: presently we discuss a buffered ion trap design that is optimal for entanglement purification,  maximising the  processing speed of a computer formed from such cells. A cell possessing tens of qubits could also embody multiple data qubits -- we do not pursue this possibility here, but it is an obvious method for reducing the number of cells required for a given computational task. In any case a useful quantum computer will require a great many basic cells, but since each cells is likely to be of sub-centimetre dimensions a machine comprised of millions of cells could fit within the space allocated to a conventional supercomputer.

To support quantum computing with the cellular paradigm, we must select an approach to achieving fault tolerance -- this will effectively set the target fidelity with which purified inter-cell operations must be performed. We opt to employ a  surface code, first introduced by Kitaev \cite{Kitaev97,Kitaev03fault-tolerantquantum}, because of its high thresholds and local structure \cite{DENNIS_topolQuantumMem}. The approach involves repeatedly measuring certain {\em stabilizers} -- these correspond to simple parity measurements on groups of qubits, i.e. we need to learn whether the total number of 1's in the group is odd or even. The basic repeating cycle of the computer involves alternating patterns of parity checks separated by Hadamard rotations to switch between the $x$ and $z$ basis.  Remarkably, this simple principle allows for far more than merely protecting quantum information from errors: certain operations between encoded logical qubits can be performed merely by altering the patterns of parity measurements~\cite{Kitaev03fault-tolerantquantum,HORSMAN_Lattice_surgery} and together with a technique such as magic state distillation \cite{BRAVYI_magicStates}, all the operations required for universal quantum computation can be performed this way. 

\begin{figure}
\includegraphics[width=8.5cm]{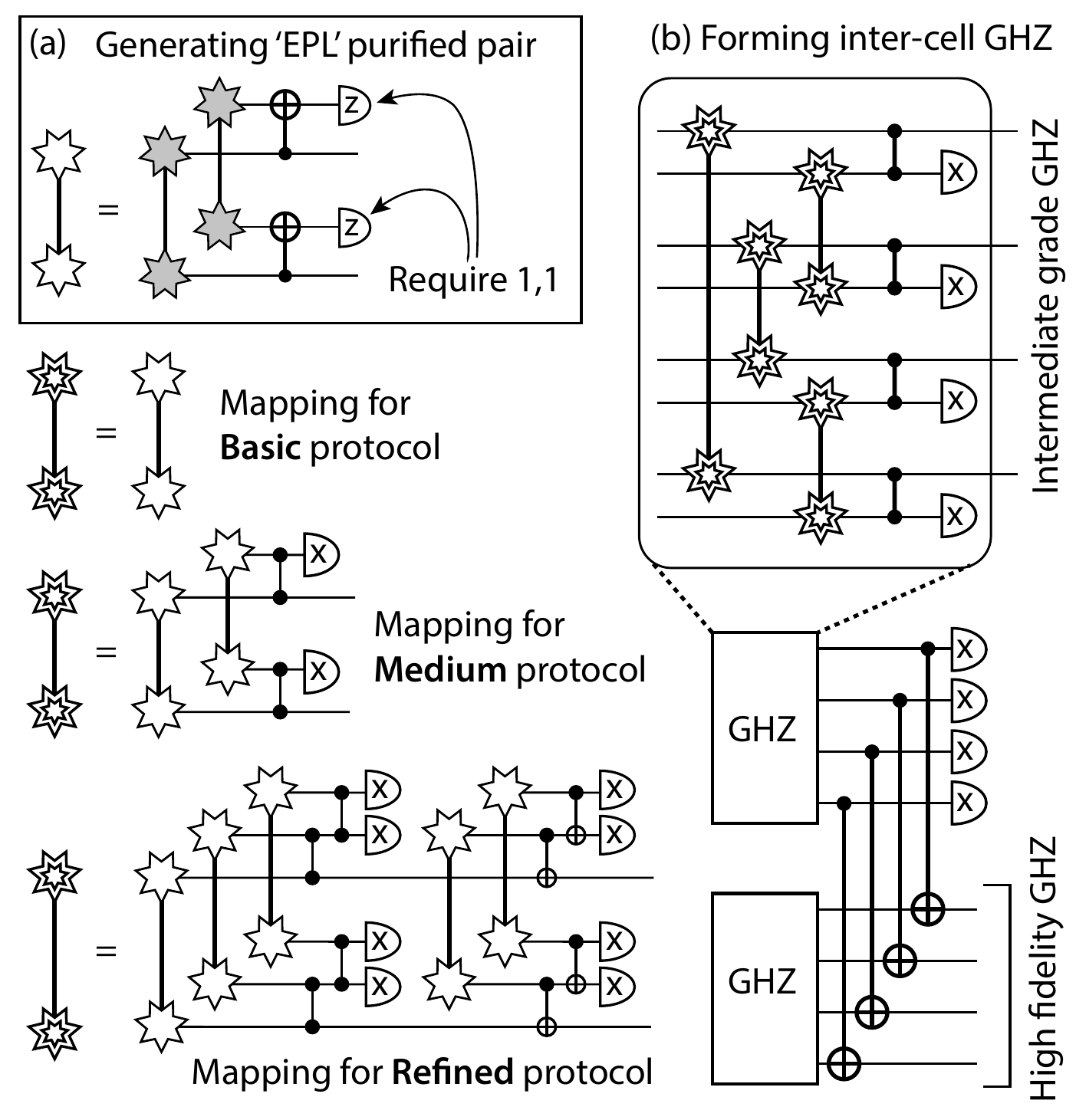}
\caption{\label{fig:epl-ghz-distillation}\textbf{Distilling a GHZ state with
EPL generated entanglement. }(a) The symbols with grey-shaded stars represent the highly 
mixed Bell states $\rho_{\rm raw}$ or  $\rho_{\rm raw}^\prime$ obtained when a single detector `click' 
is seen. Following the the Extreme Photon Loss protocol~\cite{CAMPBELL_extremePhotonLoss} we can combine two such pairs and measure one out, thus producing a single greatly improved pair. This process is represented by a symbol with 
an open star. (b) Circuit diagram showing the adapted GHZ distillation process for entanglement
 generated using the EPL protocol. Two GHZ states are produced and one is used to make a 4-qubit parity
projection onto the other. The three different protocols, Basic, Medium and Refined are shown. }
\end{figure}

In a monolithic 2D device, there is a natural layout for the physical qubits such that nearest neighbour interactions suffice to efficiently perform the parity evaluation (see Figure \ref{fig:Monolithic vs network} left).  But how should one find the parity of four data qubits if they are instead incorporated into four different cells? The solution used in Ref.~\cite{nickerson_topological} is to employ ancilla qubits within the cells, interacting them with one another across the network so as to build up a four-qubit GHZ state with one qubit in each cell (see Figure \ref{fig:Monolithic vs network} right). Given this GHZ state it is trivial to deterministically find the parity of the data qubits, as shown in the figure. The challenge is to efficiently make a high fidelity GHZ in an efficient manner. We now specify the protocols that we have developed (Fig.~\ref{fig:epl-ghz-distillation}) and we establish their performance in terms of the fault tolerance thresholds. The full process of deriving these performance figures is fairly involved, and is described in Appendix \ref{sec:simulation_methods}.

In order to minimise the impact of photon loss on entanglement generation rates (and so ultimately maximise the computer's clock speed) we must optimise the mechanism by which `raw' entanglement is achieved between cells. Typical schemes for optical entanglement generation, such as the Barrett-Kok method \cite{BARRETT_KOK} are based on heralding by two photons so that the rate of successful entanglement has a quadratic dependence on the probability that a photon avoids loss, $R_{BK}\sim\frac{1}{2}\left(1-p_{\rm loss}\right)^{2}$. Here $p_{\rm loss}$ is the
probability that an emitted photon fails to `make it' through the system and yield a detector click, whether due to loss or detector failure.
The anti-bunching scheme employed by Monroe's group in \cite{monroeNew}  has a similar form $R_{M}\sim\frac{1}{4}\left(1-p_{\rm loss}\right)^{2}$. In both cases the quadratic dependence is punishing when losses are severe and only rare photons are captured and detected. The use of cavities to enhance matter light coupling may eventually allow more sophisticated entanglement channels (as recently demonstrated~\cite{ReisererNature2014,steinerArxiv}) but here we assume that cavities are not employed, and therefore we must minimise the 
impact of loss. We adapt a scheme of Campbell and Benjamin \cite{CAMPBELL_extremePhotonLoss} called the `Extreme Photon Loss' protocol (EPL), which requires one additional qubit at each site and results in success rate $R_{{\rm EPL}}\sim\frac{1}{8}\left(1-p_{\rm loss}\right)$, i.e. linear in the photon loss rate (the precise pre-factor depends on a parameter in the scheme).

In common with Barrett and Kok's approach, this scheme requires a system capable of conditionally emitting a photon depending on its state (see lower left panel in Fig.~\ref{fig:paradigm}). Two optically active `broker' qubits~\cite{brokerClient}, each in a separate cell, are initialised to a state $\sqrt{p_{0}}|0\rangle+\sqrt{p_{1}}|1\rangle$ and optically excited causing the $|1\rangle$ component to emit a photon as it immediately decays from a short-lived excited state. Any such photons emitted pass through a beam splitter before impinging on photon detectors. The EPL protocol assumes that useful entanglement is heralded by the detection of a single photon `click'. In the absence of photon loss or noise, this would produce an odd parity Bell state, say $|\psi\rangle\equiv(\ket{01}+\ket{10})/\sqrt{2}$. In reality there are various sources of error in this process. The first is the consequence of general imperfections in the preparation and manipulation of the qubits, which we model by mixing the ideal Bell state with the identity, leading to

\begin{equation}
\rho_{\rm imperfect}=\left(1-p_{n}\right)\ket{\psi}\bra{\psi}+\frac{p_{n}}{3}\sum_{i=1,2,3}\ket{\phi_{i}}\bra{\phi_i}   \label{eq:werner_form}
\end{equation}
where the $\phi_{i}$ are the other three Bell states. We note that noise model is actually quite general: any state can be `twirled' into this form using local operations, which (being relatively high fidelity) will not significantly degrade the entanglement. However if the imperfections in the system lead to a state with biased noise, as for example if phase noise dominates, then in fact this bias may be advantageous~\cite{EarlPRA}; it is typically most difficult to purify structureless `white' noise of the form assumed here. 

Now in addition to this general noise we have the specific problems of photon loss and dark counts. With photon loss the primary issue is that when we see a single `click', as required by the protocol, it may be that that in fact two photons were emitted, one from each qubit, but one photon was lost and we thus incorrectly heralded a success. In that event the eventual broker state is $\ket{11}$. Because of this possibly, our state is 

\begin{equation}
\rho_{\rm raw}=\left(1-r\right)\rho_{\rm imperfect}+r|11\rangle\langle11| \label{eqnRealRaw}
\end{equation}
where $r=p_{\rm loss}/(p_{1}^{-1}-1+p_{\rm loss})$. In our simulations we assume that $p_{\rm loss}$ is very severe -- it approaches unity and therefore $r\approx p_1$. Thus $\rho_{\rm raw}$ is highly mixed: its two terms will have comparable weight. 

Having thus accounted for photon loss, it remains to assess the impact of dark counts. Our protocol proves to be quite robust versus this issue; a full analysis is presented in Appendix~\ref{sec:DarkCounts}. We show that the key parameter is $d=p_{dc}/(1-p_{\rm loss})$ where $p_{dc}$ is the probability that a given detector registers a dark count in the detection window of a single entanglement attempt. Provided that $d\lesssim 10^{-2}$ then to a good approximation dark counts simply increase the network infidelity; for example if we set $r=1/2$ then finite dark counts result in $p_n\rightarrow p_n+3d$.

Given that we have seen a detector `click' and so heralded the existence of $\rho_{\rm raw}$, we now store this state and proceed to create another instance of it. Note that there will typically be many heralded failures, i.e. instances where no detector `click' is reported, before another success is seen. When that success occurs we again have a $\rho_{\rm raw}$, except that we apply an additional random phase shift to account for the fact that a substantial time may pass between creation of the two pairs, so that a finite unknown phase drift in the network may have occurred. This case be can modelled by saying that a $\pi$ phase shift has occurred with probability $p_{\rm drift}$, as follows

\begin{equation}
\rho_{\rm raw}'=\left(1-p_{\rm drift}\right)\rho_{\rm raw}+p_{\rm drift}Z_{1}\rho_{\rm raw}Z_{1}. \label{eqnRealRawPrime}
\end{equation}

Note that apart from this possible drift {\em between} the two heralded successes, the approach is otherwise interferometrically stable: an unknown phase shift that is acquired by both $\rho_{\rm raw}$ and $\rho_{\rm raw}'$ will cancel out in the next step, see Appendix~\ref{sec:EPLexplain}. This step proceeds as shown in Fig.~\ref{fig:epl-ghz-distillation}. Within each of the two cells a local control-{\small NOT}  operation is performed; it is controlled by the broker associated with $\rho_{\rm raw}$ and targets the broker associated with $\rho_{\rm raw}'$. The brokers associated with $\rho_{\rm raw}'$ are then separately measured in the $z$ basis. As explained in Appendix \ref{sec:EPLexplain}, the measurement outcome $1,1$ is inconsistent with {\em either} of the two pairs $\rho_{\rm raw}$ or $\rho_{\rm raw}'$ having been originally in the $\ket{11}$ state; thus if that outcome is seen, the $\ket{11}\bra{11}$ component of the surviving entangled pair is removed. A convenient feature of this protocol is that the desired measurement outcome, $|11\rangle$, can be made to correspond to `bright' states of the matter qubits which have a higher measurement fidelity than their `dark' counterpart in several optical systems.

\begin{figure*}
\centering{}
\includegraphics[width=2.06\columnwidth]{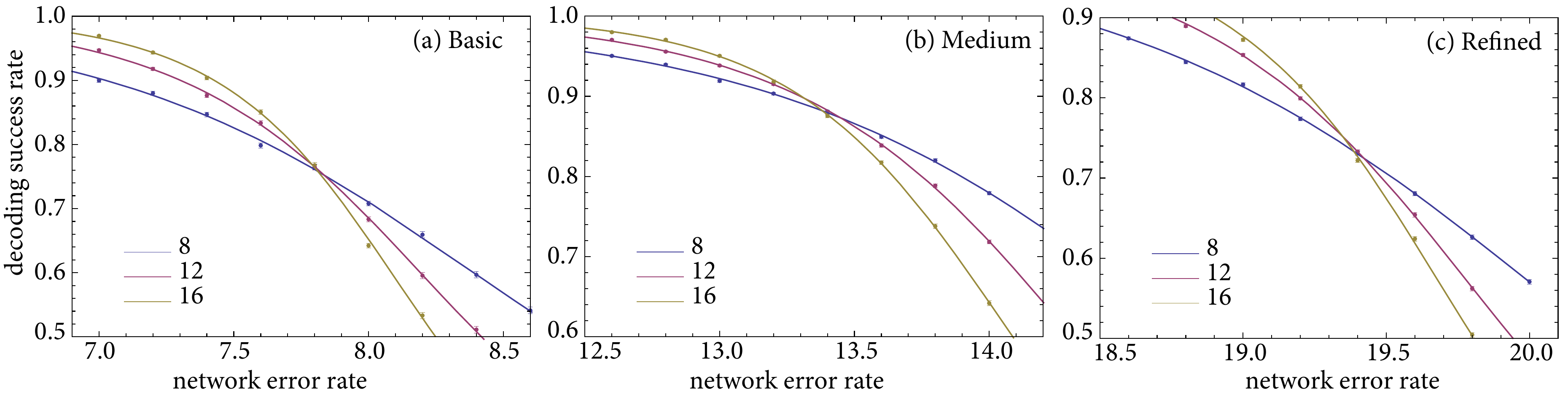}
\caption{\label{fig:Results}
  Results of threshold calculations for the three protocols considered, a) Basic   b) Medium and  c) Refined. 
  The logical error rate in the toric code is calculated for varying values of the network error rate, $p_{n}$. The cells' internal error 
  rates are taken to be the best currently demonstrated in an ion trap system:  a two qubit gate error rate of 0.1\% and a 
  measurement error rate of 0.05\%. Infidelity in the single qubit rotations is taken to be at least one order of magnitude less than these errors. We select $p_1=\frac{1}{4}\ \Rightarrow\ r\approx\frac{1}{4}$ and we take the phase shift in the EPL entanglement generation to be $p_{\rm drift}$, of 1\%. Details of the error model can be found in Appendix \ref{sec:simulation_methods}. The three curves on each plot denote the results for 
  increasing lattice sizes, where $L=8$, $12$, and $16$ (as defined in Appendix \ref{sub:threshold_calculation}). The threshold is defined as the intersection of these curves from which we find  Basic has a threshold of $p_n=7.7\%$,  Medium a threshold of  $p_n=13.3\%$ 
  and  Refined a threshold of $p_n=19.4\%$. In this paper we employ the toric surface code where the boundaries of the network are periodic; we have confirmed that the performance of the alternative planar variant of the code is very similar, see Appendix \ref{sec:planar code results}.
  }
\end{figure*}

Of course the local intra-cell operations and measurements must themselves be treated as noisy (see Appendix \ref{sec:simulation_methods} for the noise model). Given that they are reasonably high fidelity then the result of a successful ``$1,1$'' outcome is the Bell state $\rho_{{\rm EPL}}$ 
which, while still imperfect, is far higher fidelity than the parent states $\rho_{\rm raw}$ and $\rho_{\rm raw}'$. We then take these EPL-derived Bell pairs $\rho_{{\rm EPL}}$ as the basic resource for our GHZ creation. (Optionally we could use the EPL protocol to perform some, or all of, the parity projections involved in creating the GHZ, however this possibility is not explored here.) We introduce three new purification protocols with varying time-versus-fidelity tradeoffs. These are depicted in Figure \ref{fig:epl-ghz-distillation}(b). The Basic protocol is fast but tolerates only a limited network error rate. The Refined protocol carries out several rounds of entanglement distillation making it much more robust against noise, but also quite time consuming. The Medium option sits between these two extremes. 

Having established our procedure for generating shared GHZ states across the network links, we proceed to determine the performance of the quantum machine by simulating and tracking errors. This is an intensive numerical process benefiting from the use of a cluster-scale computer facility. The process is detailed in Appendix~\ref{sec:simulation_methods}, here we summarise it: For a given set of local error rates, we pick a network error rate $p_n$, a network size (number of cells) and we simulate a large number of stabilizer cycles of the computer. At the end of this simulation we inspect the state to determine whether the logical qubit was corrupted, a simple `yes'/`no' outcome. 
We repeat this numerical experiment many thousands of times (typically $3\times 10^4$) to determine the probability that logically encoded qubits will survive these stabilizer cycles without error; this produces one data point for  Figure~\ref{fig:Results}. This process is now repeated with a different network {\em size} -- if the larger network has a lower logical qubit error rate, we deem that the surface code is operating successfully and therefore our chosen network error rate was within the threshold for fault tolerance. The analysis is then repeated for a different levels of network noise in order to determine the threshold precisely.

The results of these calculations are shown in Figure~\ref{fig:Results}, from which we find that the Basic protocol leads to a threshold of $7.7\%$,  the Medium complexity protocol has a threshold of  $13.3\%$  while the most aggressive protocol, Refined, is able to tolerate very high network noise of up to $19.4\%$; note in each case we allowed for an additional $1\%$ phase drift between the two rounds of EPL.

The question of how these protocols behave when well-below-threshold (i.e. the regime where a device would realistically operate) requires a different approach to the Monte Carlo simulations performed here, as considered in several recent works~\cite{Bravyi_RareEvents,Watson_overheads}. While this is beyond the scope of the present paper, we have noted that using the Medium protocol at half the threshold network error ($7\%$) with a lattice size of $L=16$ yields a logical error rate per $L$ stabilizer rounds  of fewer than one in a million.

These simulations establish the tolerable levels of error, which are comparable to (but better than) 
our earlier paper Ref.~\cite{nickerson_topological}. However because this new approach is founded on the EPL protocol for entanglement generation, the overall time needed to perform a stabilizer -- and hence, the fundamental `clock cycle' of the quantum computer -- will be much faster than in prior schemes. 
\begin{figure}
\includegraphics[width=\columnwidth]{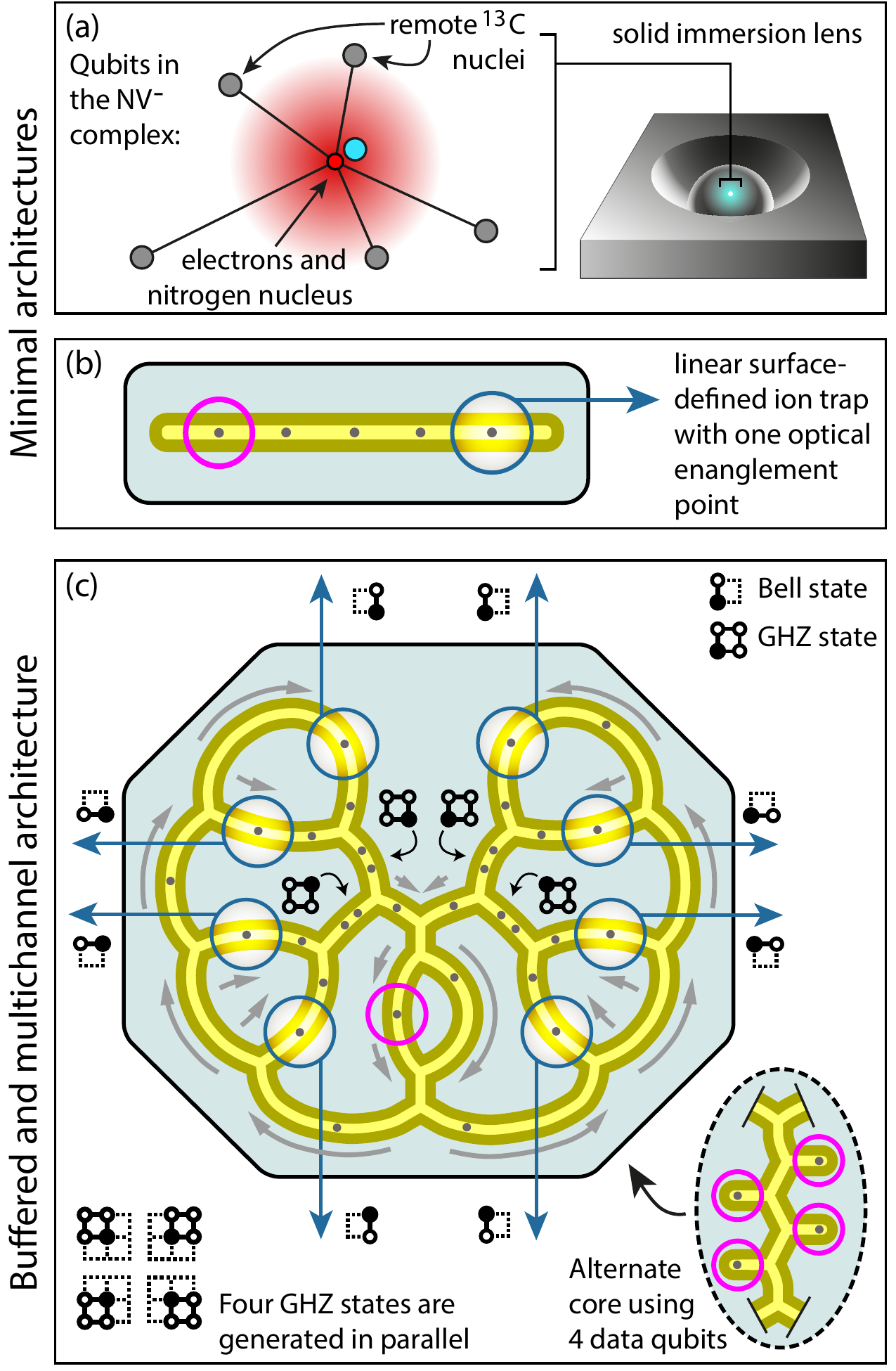}
\caption{\label{fig:trap_designs} Example architectures relevant to the cellular network paradigm. In (a) and (b) there are $5$ qubits available, the minimum required by the protocol: (a) An NV centre with several $^{13}C$ atoms within range of the core constitutes a 5-qubit cell with one qubit (the electron spin) coupling to an optical channel. A simple ion trap (b) need only have 5 ions, but a more complex architecture (c) offers the advantage of temporarily storing, or `buffering' the incomplete GHZ states. In this illustration the eight independent entanglement sites further enhance the GHZ generation rate and thus increase the `clock speed' of the computer; this also obviates the need for optical switching, as illustrated in Fig.~\ref{fig:3D_rendering}. The small square symbols indicate the generation of Bell pairs between cells, and the subsequent synthesis of GHZ states out of those Bell pairs. A filled circle indicates an ion in this cell, open circles are ions in neighbouring traps; this may be more apparent from multi-cell schematic Fig.~\ref{fig:bufferedNetwork} in the Appendices.}
\end{figure}

The achievable computer speed depends on the kind of cell architecture that we have available, see Figure~\ref{fig:trap_designs}. An obvious advantage is to have multiple entanglement channels connected to each cell; the example in Fig.~\ref{fig:trap_designs}(c) has eight channels (two to each of its four neighbouring cells, see also Fig.~\ref{fig:bufferedNetwork}). There is also a second, independent characteristic of the architecture which we analyse by distinguishing two limiting cases, the `minimal' architecture and the `buffered' architecture. In a minimal system there are just enough qubits to perform our protocols. There will then be uncertainty as to how long it takes to complete a given stabilizer measurement (since the protocols are probabilistic) and this necessitates a delay for synchronisation, see Appendix~\ref{sub:probabilistic_stabs}. In contrast, a buffered architecture has additional internal storage allowing us `queue' our qubits and smooth out the timing irregularities, so avoiding the difficulty in synchronisation. Table~\ref{tab:times} summarises the time cost to perform a high fidelity stabilizer measurement on four data qubits. It is quantified in terms of $T_0$, the time to produce a single basic EPL Bell pair, i.e. the state $\rho_{\rm EPL}$. 

\begin{table}[!b]
\begin{tabular}{| p{1.5cm} | p{2cm} | p{2cm} | p{2.5cm} |}
  \hline
  ~ & Threshold & \multicolumn{2}{c|}{Time to make GHZ (units of $T_0$)}  \\ \cline{3-4}
Protocol & error rate $\:\:\:$  ($p_n+p_{\rm drift}$) & minimal architecture & buffered architecture  \\
  \hline \hline
  Basic 	& $7.7\%+1\%$ 		& $22$	 & $5.2$ \\
  Medium	 & $13.3\%+1\%$		 & $47$ 	& $12.2$ \\
  Refined	& $19.4\%+1\%$ 		& $102$ 	& $31.6$ \\
  \hline
\end{tabular}
\caption{\label{tab:times}
The threshold of tolerable network error rates for each of the three distillation protocols considered, and the time cost for making a complete high fidelity 4-qubit GHZ assuming we operate well under threshold (3\%, 5\% and 7\% for the three protocols respectively). Such a GHZ state enables a stabilizer measurement. The distinction between minimal and buffered architectures is illustrated in Fig.~\ref{fig:trap_designs}.}
\end{table}

Given all of these contributing factors we can now estimate an achievable rate for the `clock cycle' of our computer. We will neglect the time for local gates and measurements, thus our estimate will be accurate only if such gates are performed on the scale of microseconds. This appears achievable but we note that in established experiments the highest fidelities are seen for longer gate times (see e.g.~\cite{IonBest2Q}). One cycle of a surface code quantum computer corresponds to a set of parity measurements over all its data qubits (either in the $x$-basis or the $z$-basis, alternatingly). Suppose that the cells in our machine correspond to the design in Fig.~\ref{fig:trap_designs}(c), or a superconducting qubit device of equivalent complexity. Assume that each cell-cell link is an entanglement channel which is realistic with today's technology: the entanglement attempt rate is $2$~MHz and the end-to-end photon detection probability is only $2\%$. We select $p_1=\frac{1}{4}$ and find that the average time cost for an entanglement channel to create one Bell resource ($\rho_{\rm EPL}$) is $T_0=0.27$ ms. Now further assume that we have opted for the Medium purification protocol because we have network noise at level of $5\%$ (well within Medium's threshold of $13.3\%$). According to Table~\ref{tab:times} a single-channel cell will require time $12.1 T_0$ to create one high fidelity GHZ state. Our cells have $8$ channels which together generate such GHZs at a rate of $2.5$~kHz, however two GHZ states per cell are consumed in making a complete set of stabilizer measurements (either $x$-basis or $z$-basis) as explained in Appendix~\ref{sub:scheduling}. Therefore our overall clock rate is $1.2$~kHz.

Higher rates could be achieved simply by introducing more entanglement channels (the branched design in Fig.~\ref{fig:trap_designs}c obviously generalises from $8$ channels to $2^N$). This would be consistent with ideas in the MUSIQC project~\cite{MUSIQC}. Alternatively if we look to the medium term future and assume that the use of integrated cavities~\cite{steinerArxiv} (or other advances) can reduce the photon loss rate to $\sim50\%$, and that the network noise can be taken well below the $7.5\%$ threshold of our Basic protocol, then the same device design in Fig.~\ref{fig:trap_designs}c should begin to approach megahertz rates for stabilizer measurement. At this point the local gate speeds may be the limiting factor.

\begin{figure}
\includegraphics[width=0.99\columnwidth]{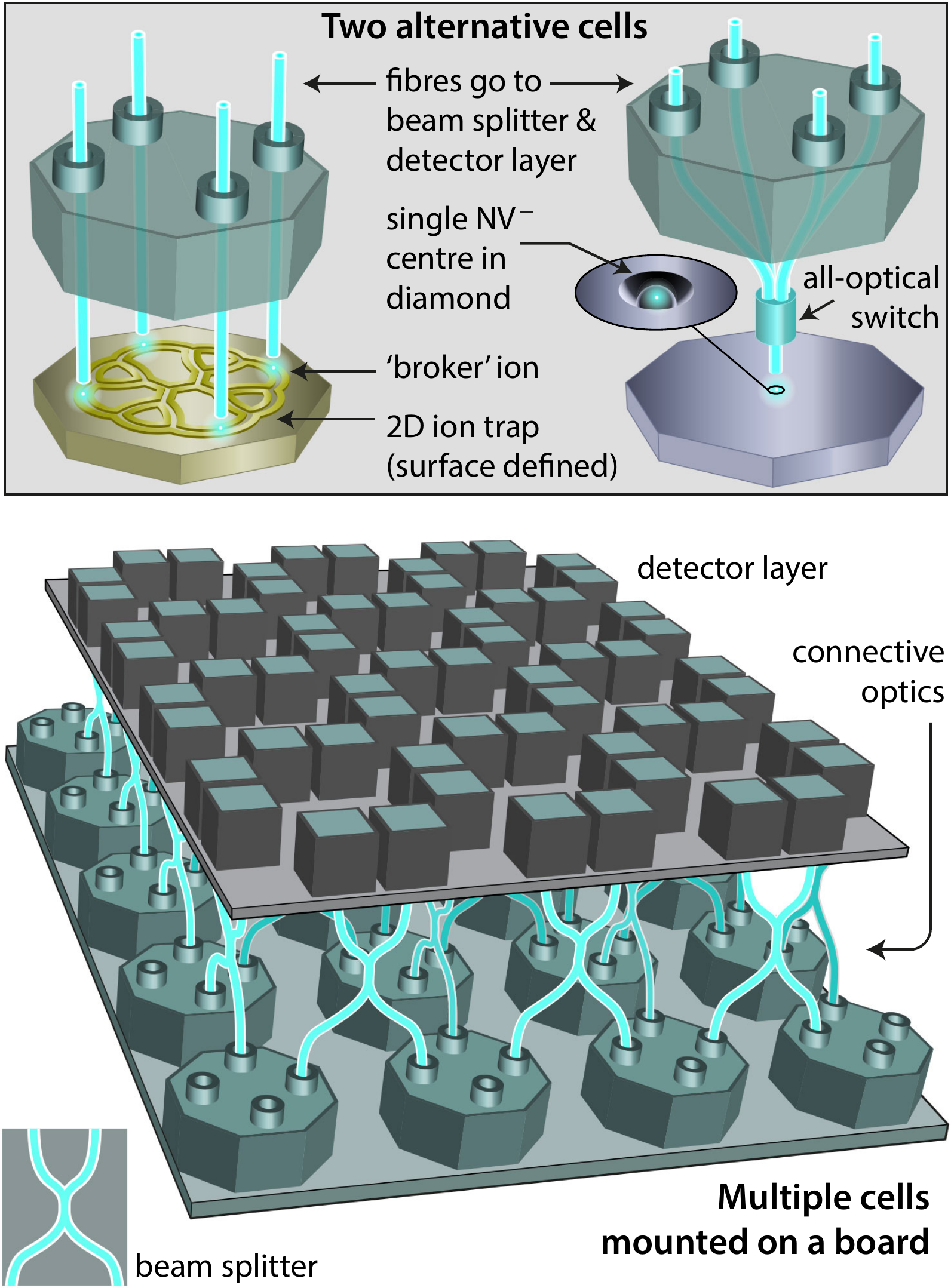}
\caption{\label{fig:3D_rendering} Illustration of the cellular architecture in 3D. See also the detailed schematics Figs.~\ref{fig:minimalNetwork} and \ref{fig:bufferedNetwork} in the Appendices.}
\end{figure}

In conclusion, we have considered an architecture for quantum technologies which is motivated by the recent achievements in ion traps, superconducting devices and NV centres. We consider small quantum `cells' comprised of a few ($5$-to-$50$) qubits under high fidelity deterministic control, together with inter-cell network links which are both noisy and lossy and therefore non-deterministic. This architecture is relevant to communication when the links are long (e.g. kilometres) but our focus here has been on quantum computing using a large number of such cells with short (e.g. centimetre) bridges. We find that the exquisitely high levels of control recently achieved in small systems can enable very compact and efficient entanglement purification, allowing one to use relatively poor photonic links between cells. We consider three different purification protocols  and derive the corresponding thresholds for fault tolerant quantum computing, finding that threshold network fidelity can be as low as $80\%$. Moreover we study the time cost of the purification process, and thus the time to evaluate a set of stabilizer measurements across the network -- effectively, the `clock speed' of the quantum computer. We relate this speed to the complexity of the cell. Given cells that are sufficiently complex to incorporate parallel operations and buffering (temporary storage) we find that even the highly lossy links that are realistic today should support kilohertz rate, freely scalable quantum computing. 

We acknowledge very useful conversations with Chris Balance, Earl Campbell, Austin Fowler, Ronald Hanson, Tom Harty, Winfried Hensinger, and David Lucas. Computing facilities were provided by the Oxford's ARC service, the Imperial College High Performance Computing Service, and the Dieter Jaksch group. SCB acknowledges EPSRC platform grant EP/J015067/1. This material is based on research funded in part by the Singapore National Research Foundation under NRF Award NRF-NRFF2013-01.

\newpage


\bigskip

\centerline{ \bf \Large Appendices}
\section{Explanation of the EPL Method\label{sec:EPLexplain}}

In the following we describe the EPL protocol's handling of photon loss and its inherent tolerance of systematic phase errors. We neglect all other imperfections, but of course these are accounted elsewhere in our analysis and in our simulations (c.f. Appendix~\ref{sec:ErrorModel} and \ref{sec:DarkCounts}).

For simplicity let us suppose we initialise each of our two remote `broker' qubits into the state $(\ket{0}+\ket{1})/\sqrt{2}$. (Note that in fact we need not start from an equal superposition; one selects an optimal level of excitation). The $\ket{1}$ state is optically active; we excite both brokers and route the collected light through a beam splitter prior to detection as in the standard picture. If we see a photon, then we have
\[
\rho_{\rm simple}=(1-r)\ket{\Theta}\bra{\Theta}+r\ket{11}\bra{11}. 
\]
where $\ket{\Theta}=(\ket{01}+e^{i\phi}\ket{10})/\sqrt{2}$ and $\phi$ is some phase introduced due to the optical apparatus. The ratio $r$ depends on the severity of photon loss; in the limit of loss tending to unity we find $r\rightarrow\frac{1}{2}$. An intuitive explanation of this is as follows:  the state $\ket{\Theta}$ generates only one photon whereas $\ket{11}$ generates two, however the state $\ket{\Theta}$ has twice the probability of $\ket{11}$ in the original broker-broker product state, thus these factors compensate and the two states have equal weight in the mixture. Now we store $\rho_{\rm simple}$ and attempt to create another such pair. We may fail a number of times before we again succeed. Provided that the apparatus does not suffer phase `drift' {\em between} the two successful events, then the second entangled pair will have the exact same form $\rho_{\rm simple}$. (Recall that our full analysis in the main paper does account for the possibility that such a drift occurs.) 

Having obtained two noisy entangled pairs, each of the form $\rho_{\rm simple}$, we now apply CNOT gates locally within
each cell according to the Fig.~\ref{fig:epl-ghz-distillation}(a) circuit. These CNOTs map each of the pure states within our $\rho_{\rm simple}\otimes\rho_{\rm simple}$ mixture as follows, where we take the left-side to be the controlling qubits, and the right-side as the target qubits:
\begin{eqnarray}
\left(\ket{01}+e^{i\phi}\ket{10}\right)\ \left(\ket{01}+e^{i\phi}\ket{10}\right) \rightarrow\ \ \ \ \ \ \ \ \ \ \  \ \ \ \ \ \ \ \   \ \ \ \  \label{eq:epl_terms}\\
\ \   \ \ \ket{01}\left(\ket{00}+e^{i\phi}\ket{11}\right)+e^{i\phi}\ket{10} \left(\ket{11}+e^{i\phi}\ket{00}\right)\ \     \nonumber
\end{eqnarray}

\begin{eqnarray}
\ket{11}\left(\ket{01}+e^{i\phi}\ket{10}\right)\ &\rightarrow&\ \ket{11}\left(\ket{10}+e^{i\phi}\ket{01}\right)\nonumber\\
\left(\ket{01}+e^{i\phi}\ket{10}\right)\ket{11}\ &\rightarrow&\ \ket{01}\ket{10}+e^{i\phi}\ket{10}\ket{01}\nonumber\\
\ket{11}\ket{11}\ &\rightarrow&\ \ket{11}\ket{00}\nonumber
\end{eqnarray}

The protocol then calls for the one of the qubit pairs, the right-side pair in the present notation, to be measured in the $z$-basis. Any result other than ``$11$'' is rejected. We see that only the first of the four possibilities listed above can pass this filter. Revisiting Eqn.~(\ref{eq:epl_terms}) and collecting terms we have
\begin{eqnarray}
&\ket{01}\left(\ket{00}+e^{i\phi}\ket{11}\right)+e^{i\phi}\ket{10} \left(\ket{11}+e^{i\phi}\ket{00}\right)&    \nonumber \\
=&\left(\ket{01}+e^{2i\phi}\ket{10}\right)\ket{00}+e^{i\phi}\big(\ket{01}+\ket{10}\big)\ket{11}. & \nonumber
\end{eqnarray}
Thus measuring ``$11$'' implies that the remaining qubit pair is state $(\ket{01}+\ket{10})/\sqrt{2}$ and we have eliminated both the $\ket{11}$ component due to photon loss and the unwanted phase $\phi$ (which, therefore, we need not know).  This occurs with probability $\frac{1}{2}(1-r)^2$, i.e. $\frac{1}{8}$ when $r=\frac{1}{2}$. 

In practice of course there are other sources of error, both in the network and in the local gates and measurements, as discussed in the error model below. But for all levels of noise relevant to the devices we are considering the result of this process is to generate a state $\rho_{\rm EPL}$ which is far higher fidelity than the two parent states.

\section{Planar Versus Toric Network Topologies \label{sec:planar code results}}

The main paper presents results for the toric code. In this version of the surface code the boundaries of the surface wrap to form a torus. While this would be difficult to realise if we were employing a monolithic structure (c.f. Fig.~\ref{fig:Monolithic vs network}~left), in a network paradigm there is no in-principle difficulty. However, it might be that there are reasons to prefer to layout the network in 2D and maintain all links of the same physical length - in this case one would adopt the planar version of the surface code. Our threshold finding numerical programs can simulate either the toric or the planar variant. We find that while the toric code exhibits a slightly sharper threshold because it has no `edge effect', in fact the value of a threshold obtained from the two approaches (given all other factors are held the same) varies only slightly.

\section{Simulation Methods\label{sec:simulation_methods}} 

Here we give an overview of the methods used in calculating threshold values for our system. The approach can be divided into two distinct sections: In the first we derive superoperators representing the net effect on the data qubits of our stabilizer measurement protocols with all their various errors. In  the second part we use a classical algorithm to track the effect of these superoperators as we simulate a surface code embodying logical qubits.

\begin{figure*}
\centering{}\includegraphics[height=7cm]{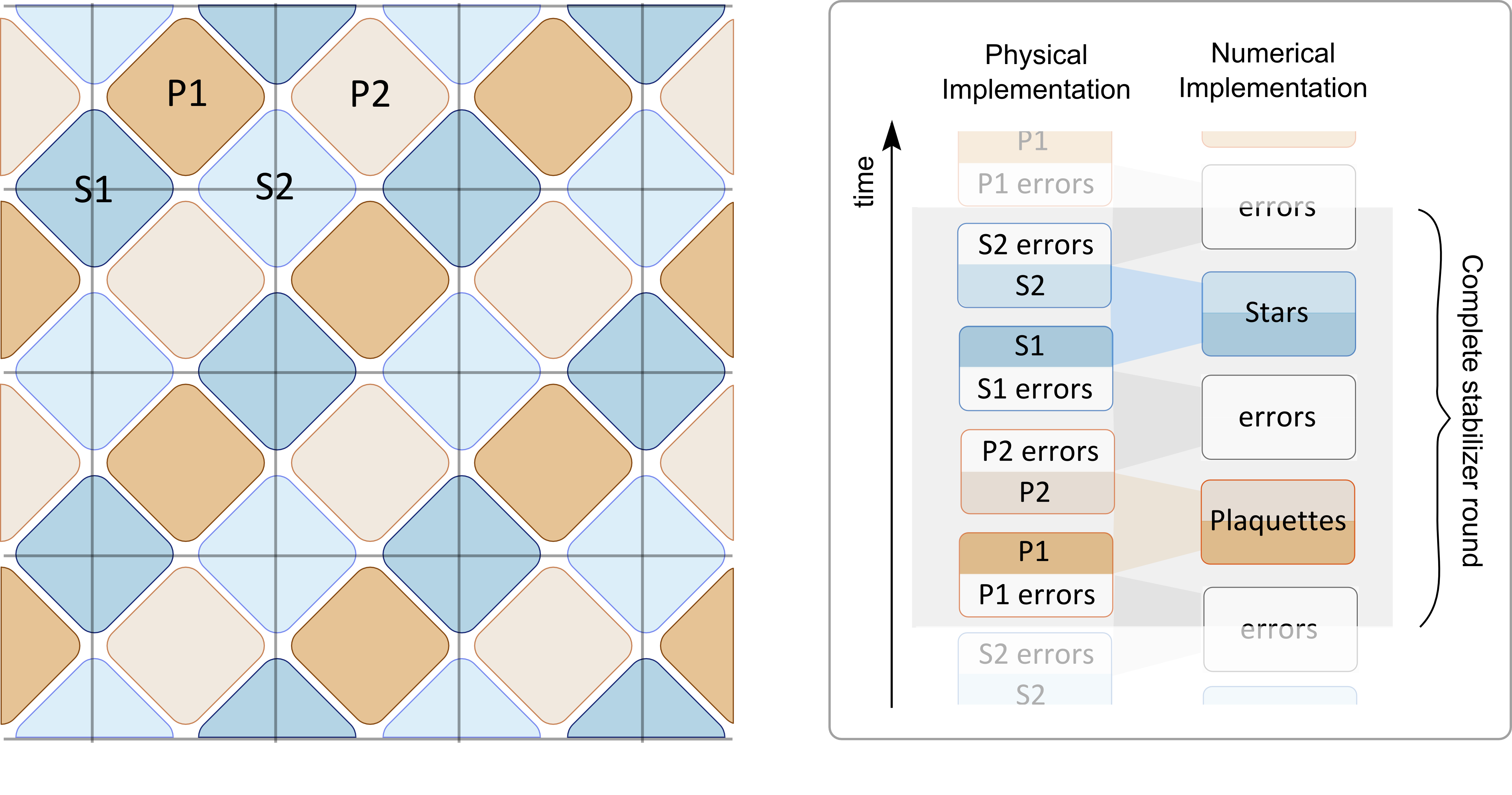}
\caption{\label{fig:scheduling}\textbf{Scheduling stabilizer measurements}
  Stabilizer measurements are split into four rounds, two of each
type plaquettes (orange) and stars (blue). The underlying lattice
and qubit structure is shown here in black for a planar code of lattice
dimension $L=4$. Decomposition of the stabilizer measurement procedure
allows each round to be described as a round of perfect measurements
either followed or preceded by errors. The stabilizer implementation
including these errors is shown on the right. }
\end{figure*}

\subsection{Error Model\label{sec:ErrorModel}}

All the protocols we consider are composed of a small number of low-level basic operations, each with an associated noise model. 

 \begin{enumerate}

\item Network error model. This is described in the main paper. States $\rho_{\rm raw}$  and $\rho_{\rm raw}'$ are defined in Eqns. (\ref{eqnRealRaw}) and (\ref{eqnRealRawPrime}). They involve the network noise rate $p_n$ and the photon loss rate $p_{\rm loss}$, as well as a parameter $p_{\rm drift}$ accounting for phase drift in the entanglement channel between the creation of the two states. Dark count rate $d$ can be subsumed into the network noise $p_n$ as explained later in Appendix~\ref{sec:DarkCounts}.
\item Local (intra-cell) controlled-$Z$ and controlled-$X$ gates. For a gate error rate $p_{g}$ the noise is modeled as a perfect gate operation, which with probability $p_{g}$ is followed at random with one of the 15 non-trivial two-qubit Pauli errors $\sigma_{i}\otimes\sigma_{j}$ where $i=0,1,2,3$ and $\sigma_{0}$ is the identity. If $\rho$ represents the ideal state after gate operation this noise map can be written
\begin{eqnarray}
N_{gate}\left(\rho\right) & = & \left(1-p_{g}\right)\rho\nonumber \\
 &  & +\frac{p_{g}}{15}\sum_{(i,j)\neq(0,0)}\left(\sigma_{i}\otimes\sigma_{j}\right)\rho\left(\sigma_{i}\otimes\sigma_{j}\right)^{\dagger}
\end{eqnarray}

\item Single qubit measurement in the $X$ and $Z$ bases. Given a measurement error rate $p_{m}$ then a particular outcome of the measurement, $q\in\left\{ 0,1\right\} $ corresponds to the intended projection
$P_{q}$ applied to the state with probability $(1-p_{m})$ and the opposite projection $P_{\bar{q}}$ applied with probability $p_{m}$. This noisy projector can be written as
\begin{equation}
\mathcal{P}_{q}\left(p_{m}\right)=\left(1-p_{m}\right)|q\rangle\langle q|+p_{m}|\bar{q}\rangle\langle\bar{q}|.
\end{equation}

\end{enumerate}

\subsection{Stabilizers as superoperators}

To characterise the entire process of the stabilizer measurement we carry out a full simulation of the measurement procedure including all sources of noise and use the Choi-Jamiolkowski isomorphism \cite{Jamiolkowskii} to generate a superoperator from the result. Thus we completely describe the action of stabilizer measurement procedure with
\begin{equation}
\mathcal{S}\left(\rho\right)=\sum_{i=0}p_{i}K_{i}\rho K_{i}^{\dagger}.
\end{equation}

This probabilistic decomposition describes the operation as a series of Kraus operators, $K_{i}$, applied to the initial state with probabilities $p_{i}$, which depend on the chosen protocol, noise model and the error rates. The leading term $i=0$ will have corresponding $K_{0}$ representing the reported parity projection, and large $p_{0}$. For the protocols considered here, the other Kraus operations can be decomposed and expressed as a parity projection with additional erroneous operations applied. For example if a noisy stabilizer measurement is made which returns an `even' outcome we find $K_{0}=P_{even}$, the reported even parity projection, and $K_{1}=P_{odd}$, which implies that a perfect odd parity projection was applied, but the wrong outcome was recorded -- a `lying' stabilizer measurement. All the other $K_{i}$ can be represented as $K_{0}$ or $K_{1}$ followed by single qubit Pauli errors. This decomposition then involves two distinct types of error: lies, where an incorrect outcome is recorded, and qubit errors, where a physical error occurs on a data qubit. The probability of each combination of events can be calculated from the values of the $p_{i}$. This information on stabilizer performance then enables classical simulation of a full planar code array, and its fault tolerance threshold can be assessed.

\subsection{Scheduling stabilizer measurement\label{sub:scheduling}}

Each qubit in the body of the lattice is part of four different stabilizer groups. Therefore in a physical implementation the measurement of a full cycle of stabilizers is divided into four distinct rounds, two of plaquette measurement, and two of star measurement as shown in Figure \ref{fig:scheduling}.  For the purpose of simulation it is desirable to break down the evolution of the lattice into complete rounds of perfect plaquette or star measurement separated by rounds of errors. This can be achieved by making use of the fact that each Kraus operator can be decomposed in different but equivalent ways, namely that each $K_{i}$ can be written with the Pauli errors either preceding or following a parity projection.

\subsection{Decoding}

Decoding is performed using Kolgomorov's Blossom V implementation of Edmond's minimum weight matching algorithm \cite{edmonds1965paths,kolmogorov2009blossom} to generate a `perfect' matching between stabilizer violations. To do this the syndrome on the lattice must be formulated into a weighted graph. In the case of perfect measurements each `-1' outcome in the syndrome becomes a node of a completely connected graph, where the weight of each edge is given by the distance between the corresponding two nodes on the lattice.

Multiple rounds of stabilization are performed, producing a three dimensional syndrome cube where the third dimension represents time.  Each point where one stabilizer measurement differs from its value in the previous round gives rise to a node in the graph. Matching in the spatial dimensions of the cube correct for physical errors on the lattice, while time-like matchings correct for lying stabilizers. The rate of lie-type errors and physical errors will not generally be the same, to account for this, time-like and space-like paths are weighted differently. The ratio of these weights is chosen to optimise performance.

In the toric code error chains on the surface will always result in two stabilizer violations, one at each end of the chain. In the planar code however, if an error chain reaches an edge of the lattice only one stabilizer violation will seen. To account for this, each node in the original graph is uniquely connected to a new node located at the nearest boundary position, following the method described in \cite{WANG_planarcodesl}. This gives the possibility for each `-1' stabilizer to match to a boundary as well as any other node on the lattice itself. 

The process described above is a `vanilla' implementation of the perfect matching decoder -- there are many possibilities for optimising the decoder which are not pursued in this paper. For example, it is well understood that correlating the $X$ and $Z$ errors reveals information about $Y$ errors. Moreover there are opportunities to exploit the classical information that occurs during a stabilizer evaluation: Most importantly, in the case where we are using a simple, serial architecture with no buffering available (an NV centre based technology, for example), then we need to impose a cut-off time after which an attempt to measure a stabilizer is abandoned, see Appendix~\ref{sub:probabilistic_stabs} below. At present our decoder makes no use of the information that a given stabilizer has not been evaluated, and instead simply replaces the missing information with a copy of the previous result -- this is obviously not optimal. 

We emphasise that these limitations in our decoder do not undermine the accuracy of the simulations in this paper, since the operator of a quantum computer is free to use any classical decoder she wishes when she runs the machine. Thus the thresholds that we find should be considered a lower bound on the achievable thresholds -- a better decoder should boost the performance.

\begin{figure*}
\includegraphics[width=17cm]{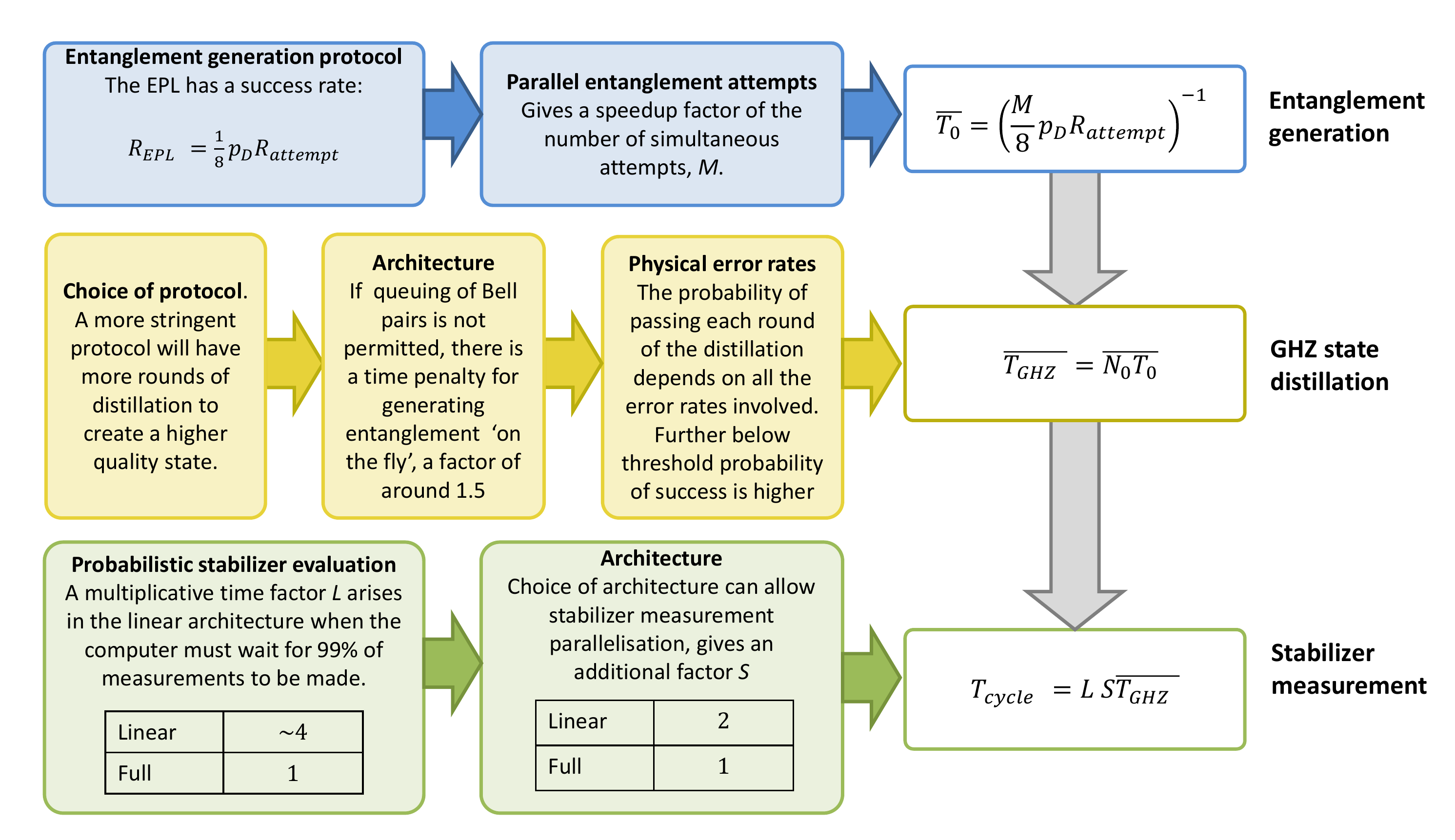}
\caption{\label{fig:running_speed}
The seven major factors contributing to the overall potential running speed of the device. 
}
\end{figure*}

\begin{figure*}
\centering{}
\includegraphics[width=17.5cm]{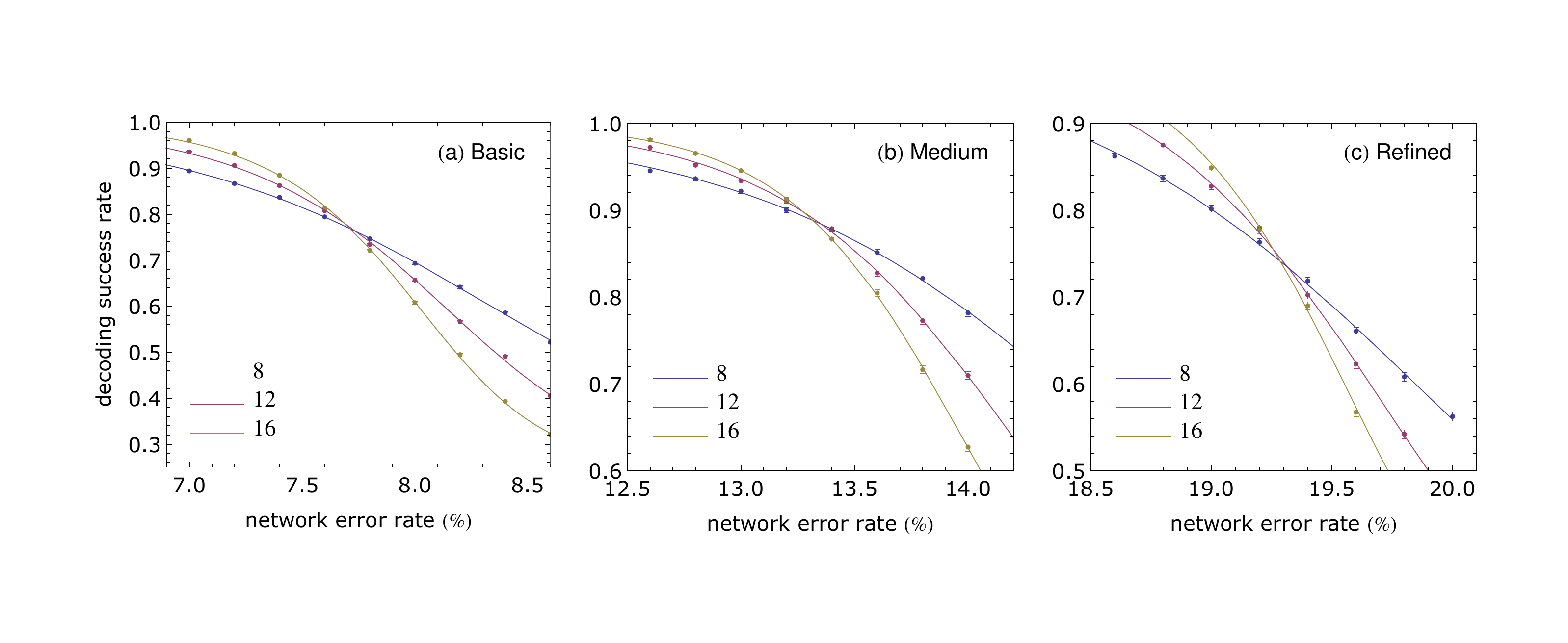}
\caption{\label{fig:missing_stabilizers}Thresholds with 1\% of stabilizer outcomes missing. a) Basic b) Medium c) Refined}
\end{figure*}

\subsection{Calculating the threshold \label{sub:threshold_calculation}}

We wish to determine the logical error rate of the code as a function of the physical error rates. To do this we perform Montecarlo simulations of a lattice under noisy stabilizer measurement. For each instance the evolution of the lattice is simulated by applying random errors drawn from the distributions specified by the derived superoperators. A total of $3L$ complete stabilizer rounds are performed, where $L$ is the lattice dimension, before a decoding attempt is made and the result analysed to test whether a logical error has occurred. For each physical error rate, the logical error rate was calculated for four different lattice sizes, $L=$ 8, 12, 16. For each data point a minimum of 30,000 instances were simulated, and error bars were calculated by treating each result as a sample drawn from a Bernoulli distribution.

If the error rate is below the threshold then increasing the lattice size will improve the performance of the code, that is the logical error rate will become  smaller. So to find the threshold we must find the point at which the curves from  the different lattice sizes intersect.   To estimate the threshold error rate we use the method described by Wang et al. in \cite{WANG_planarcodesl} to model the behaviour of the logical error rate close to the crossing. This tells us that for a large enough lattice size, $L$, the decoding failure probability his given by  

\begin{equation}
P_{fail}=\left(p-p_{th}\right)L^{\nicefrac{1}{v_{0}}}
\end{equation}

The threshold data is fitted to a quadratic function, to account for small system-size effects, and the threshold crossing value drawn from the resultant fit parameters.

\begin{equation}
P_{fail}=a+b\left(p-p_{th}\right)L^{1/\nu_{0}}+c\left(p-p_{th}\right)^{2}L^{2/\nu_{0}}
\end{equation}

\section{Computer operational speed\label{sec:running_times}}

The entanglement generation described in the previous section is only one aspect that determines the overall speed of the device. Figure \ref{fig:running_speed} shows a summary of all factors contributing to the final clock cycle of the quantum computer. The contributors are divided into three categories. Entanglement generation has already been discussed, we will now consider the factors affecting GHZ state distillation time and stabilizer measurement.

\begin{figure*}
\includegraphics[width=15cm]{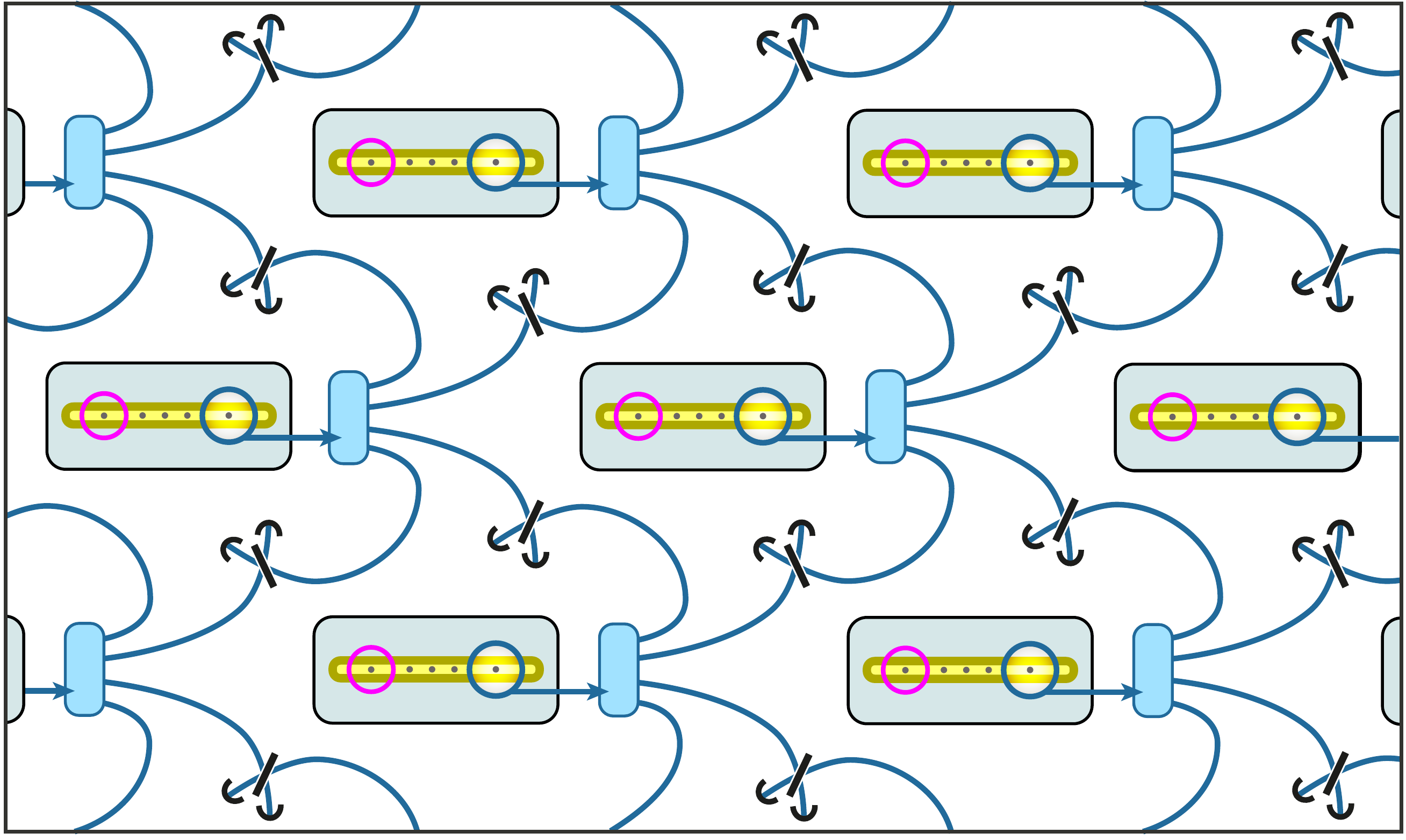}
\caption{\label{fig:minimalNetwork}
Four cells of the design shown in Fig.~\ref{fig:trap_designs}(b) with the connections achieved by optical switching. Note that the ion trap elements in this figure could equivalently be any other few-qubit, optically active system such as an NV centre in diamond. The system shown here is equivalent to that shown in \ref{fig:bufferedNetwork} which has more complex cells and dedicated (non-switched) cell-cell links; those features increase the speed but the error thresholds etc are the same. 
}
\end{figure*}

\begin{figure*}
\includegraphics[width=18cm]{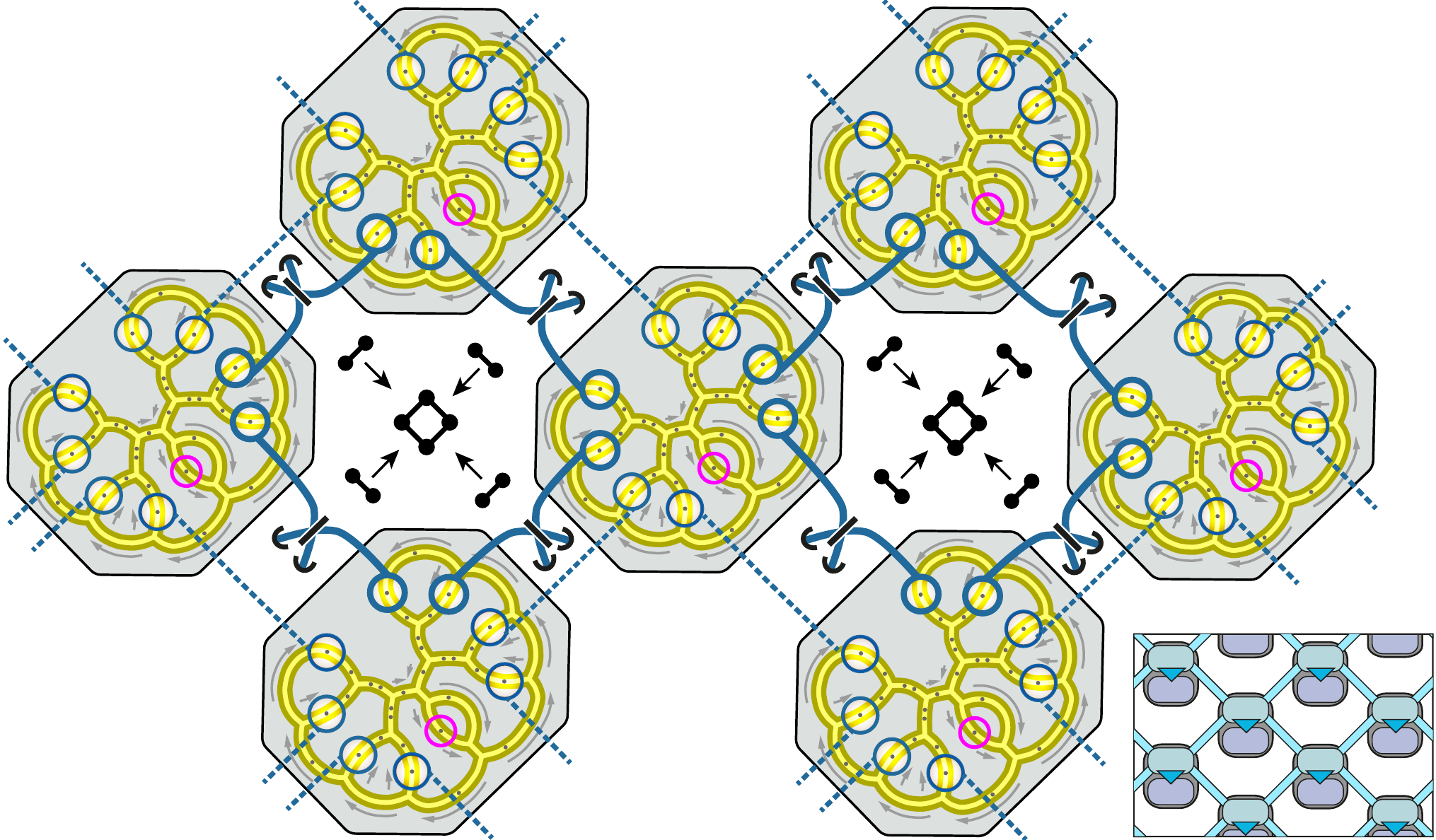}
\caption{\label{fig:bufferedNetwork}
Four cells of the design shown in Fig.~\ref{fig:trap_designs}(c) with the connections relevant to building their mutual GHZ states highlighted. The inset shows the abstract concept of of the networked computer from Fig.~\ref{fig:paradigm}; this ion trap design and the design in Fig. \ref{fig:minimalNetwork} are specific realisations.
}
\end{figure*}

\subsection{Handling probabilistic stabilizer evaluation\label{sub:probabilistic_stabs}}

The generation and distillation of entanglement using optical links between cells is a  probabilistic process, and steps must sometimes be attempted repeatedly. If the cells do not have sufficient internal complexity to queue up, or buffer, the results of the different stages of the purification, then necessarily the time taken to complete a stabilizer measurement will also probabilistic. This will be the case for the `minimal' architectures in Figs.~\ref{architectures3}(a) and (b), and in Fig.~\ref{fig:minimalNetwork}. There is a potential difficulty in performing a complete set of stabilizers (or a complete sub-set, c.f. Fig.~\ref{fig:scheduling}) over the entire computer -- should we wait until the very last stabilizer has been successfully performed, before moving to the next set? To do so would require a time cost that scales with the computer size. Fortunately this is not necessary; we can simply wait a fixed time and then abandon any stabilizers which have not yet been measured. Figure \ref{fig:missing_stabilizers} shows thresholds for the case where each stabilizer has a $1\%$ chance of not  being evaluated - thresholds change only minimally versus those in the main paper.  By thus requiring only a high proportion of stabilizers to be completed, rather than all of them, the evaluation time becomes essentially constant and independent of lattice size. We note that our GHZ-based approach to stabilizer measurement is particularly `friendly' to this process of abandoning the `slowest' $1\%$ of measurements in each round, because when we abandon an attempt to create a GHZ the data qubits in those cells have {\em not been involved} in any gate operations. Operations on data qubits only take place after successful completion of the high fidelity GHZ state.

This approach is still less than ideally efficient since on average cells will be inactive for a significant portion of their time, having finished well before the `cut off time'. Since some potential technologies for the cell, including for example NV centres, do not have the internal complexity to act as a sophisticated device such as Figure~\ref{fig:trap_designs}(c), it is interesting to ask whether there is another route higher efficiency. A possibility is asynchronous stabilizer measurement, where measurements are made not in discrete rounds but at soon as the necessary entanglement has been generated. 

\subsection{Cell design to support parallelization\label{sub:trap_architectures}}
The design of the cell can lead to a number of ways to parallelize the protocol. The full architecture shown in Figure \ref{fig:trap_designs}c is designed to exploit all of these possibilities.  At the lowest level, entanglement generation can be parallelized. If $M$ simultaneous attempts at entanglement are made then the effective rate of production is of course increased by the same factor. Further to this a cell may be able to support entanglement `queuing' such that Bell states are continuously created and stored for later use. In such an architecture entanglement can be treated as being deterministically generated at the mean rate (or slightly less, to maintain buffers). On the other hand, if  entanglement must be generated `on the fly', as required, then this must be treated as a stochastic process.  At a higher level GHZ states can be stored as and when they are created, leaving other qubits free to generate more. This removes the inefficiency discussed in the previous section, where most cells must wait for the slowest stabilizer measurements to complete. Instead stabilizers can be measured at the mean rate of GHZ distillation, providing a further level of speedup. Finally, the greater connectivity of the full architecture allows all the GHZ states required for a stabilizer round (plaquettes or stars) can be simultaneously distilled. This gives an additional speedup factor of 2 over a minimal cell where this process must be broken into two rounds as shown in Figure \ref{fig:scheduling}. 

\onecolumngrid

\section{Analysis of Dark Counts\label{sec:DarkCounts}}

We begin by considering the simple case of $r=1/2$, where it is straightforward to show that the effect of dark counts is to produce an adjusted level of network noise $p_n$. From the main paper, we have the following expressions for the case where there dark counts are neglected:
\begin{equation}
\rho_{\rm imperfect} = \left(1-p_n\right)  \ket{\psi}\bra{\psi} + \frac{p_n}{3} \left(\ket{\phi}\bra{\phi} + \ket{00}\ket{00} + \ket{11}\bra{11}\right),
\end{equation}
and
\begin{equation}
\rho_{\rm raw} = (1-r)\rho_{\rm imperfect} + r \ket{11}\bra{11}. \label{eqn:rhoRawForCD}
\end{equation}
Let us take the simple case that $r=\frac{1}{2}$, and introduce $d=p_{\rm dc}/(1-p_{\rm loss})$ as above, where $p_{\rm dc}$ is the probability that the system will experience a dark count in a given {\bf one} of the two detectors during a given attempt at entanglement.  
We will see that we can write a new expression $\rho_{\rm raw}^\prime$ which has the same form above, but where $p_n'$ and $r'$ replace the unprimed parameters and have absorbed the dark count parameter $d$. Roughly speaking, $p_n'\approx p_n+3d$.

Consider the limit of high photon loss, where almost all dark count events occur on occasions when all emitted photons have been lost. When we see a dark count on such a occasion, we wrongly conclude that we have heralded the creation of $\rho_{\rm raw}$. In fact, the state of the two optically active qubits is simply the completely mixed state, because they were prepared in an equal superposition and then (effectively) measured by the environment. Thus we can write,
\begin{eqnarray}
\rho_{\rm raw}' &=& (\rho_{\rm raw}+2d\frac{\mathbb{I}}{4})/(1+2d)\\
&=& (\frac{1}{2}\rho_{\rm imperfect}+\frac{1}{2}\ket{11}\bra{11}+\frac{d\mathbb{I}}{2})/(1+2d)\nonumber\\
&=& (1+2d)^{-1}\left[\left(\frac{1-p_n}{2}+\frac{d}{2}\right)  \ket{\psi}\bra{\psi} + \left(\frac{p_n}{6}+\frac{d}{2}\right) \left(\ket{\phi}\bra{\phi} + \ket{00}\ket{00}\right) + 
\left(\frac{p_n}{6}+\frac{d}{2}+\frac{1}{2} \right)\ket{11}\bra{11}\right]\nonumber\\
&=&(1-r')\rho_{\rm imperfect}' + r' \ket{11}\bra{11}\ \ \ \ \ \ {\rm with}\ \ \ \ \rho_{\rm imperfect}' = \left(1-p_n'\right)  \ket{\psi}\bra{\psi} + \frac{p_n'}{3} \left(\ket{\phi}\bra{\phi} + \ket{00}\ket{00} + \ket{11}\bra{11}\right)\nonumber
\end{eqnarray}
where the last line introduces
\begin{equation}
r'=\frac{1}{2+4d}\ \ \ \ \ \ {\rm and}\ \ \ \ \ \ \ p_n'=\frac{p_n+3d}{1+4d}.
\end{equation}

\begin{figure}
\includegraphics[width=0.78\columnwidth]{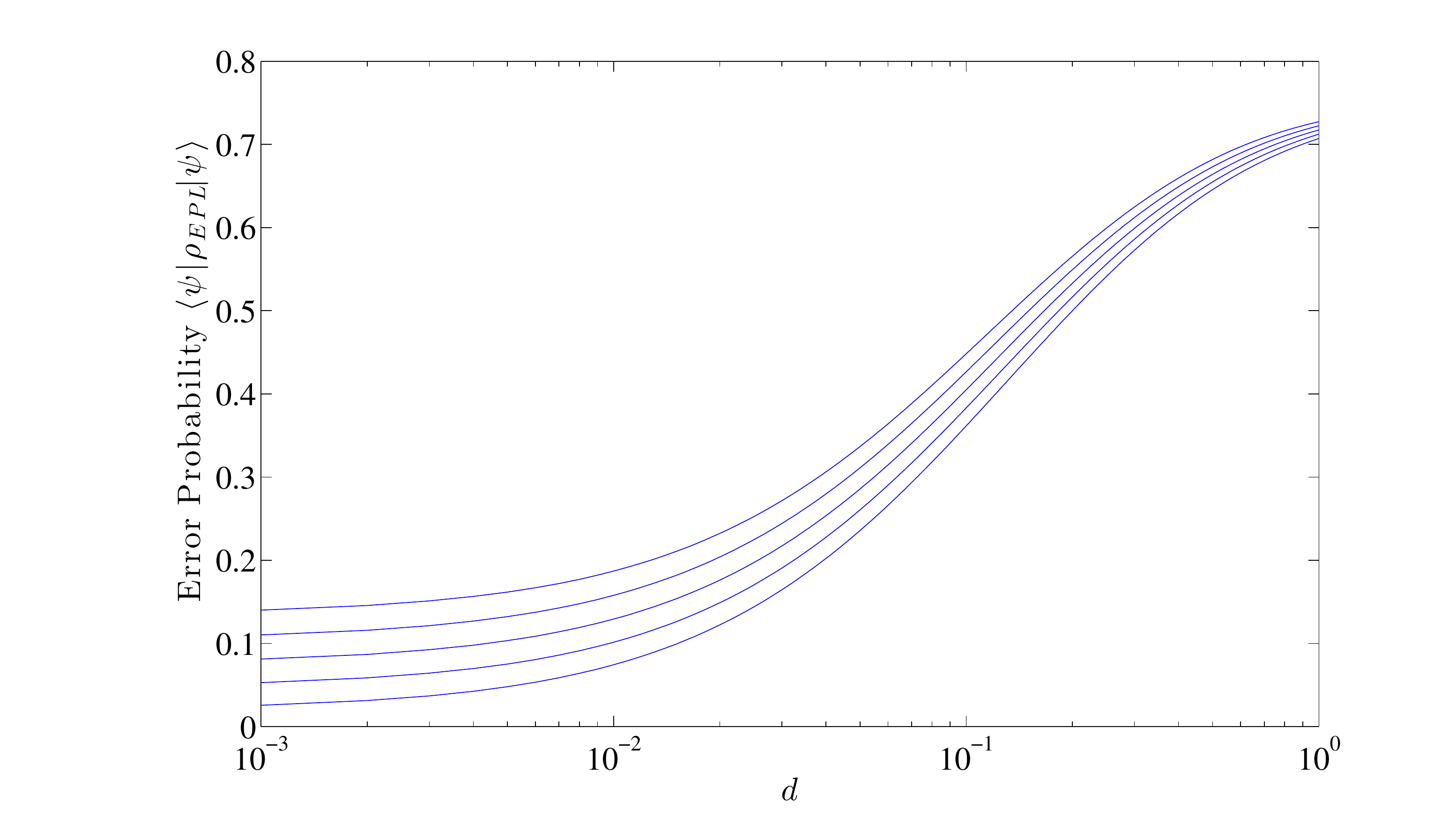}
\caption{Total error probability. Here we have taken $r=\frac{1}{4}$ and $p_{\rm drift}=0.01$. The five lines represent the total error probability corresponding  from bottom to top to preparation error probabilities $p_n = \{0,0.025,0.05,0.075,0.1\}$. The error probability is close to the error rate without dark counts while $d$ is below approximately 0.01, but increases rapidly there after, passing through the distillation threshold even for the case $p_n = 0$. \label{fig:errorprob}}
\end{figure}

We now present a more general analysis of dark counts for arbitrary $r$, again demonstrating that they serve to additively increase the effective network error $p_n$. To accomplish this, we first note that $\rho_{\rm raw}$ (defined, as in the main paper, as the state heralded by a single click, with dark counts assumed impossible) is always diagonal in the basis $\{\ket{\psi},\ket{\phi},\ket{00},\ket{11}\}$, where $\ket{\phi}$ is the antisymmetric state, see Eqn.~\ref{eqn:rhoRawForCD}. If a non-zero dark-count probability $p_{\rm dc}$ is taken into account, then the state of the system (given it has been {\em post selected} due to a single detector clicking as required) will be altered due to three additional ways in which that single click can be produced. The first is that the state $\ket{00}$ may survive post-selection due to a single dark count occuring, and this occurs with an absolute probability $p_{00} = 2 p_{\rm dc} (1-p_{\rm dc}) (1-p_1)^2$. The second way in which a dark count can lead to an effect on the post-selected state is that both a single photon loss and a single dark count can occur, resulting in the state $\frac{1}{2} \left(\ket{\psi}\bra{\psi} + \ket{\phi}\bra{\phi}\right)$ with absolute probability $p_{\psi + \phi} = 4 p_{\rm dc} (1-p_{\rm dc}) p_{\rm loss} p_1 (1-p_1)$. Finally, a combination of both the loss of two photons and a single dark count can lead to the erroneous inclusion of the $\ket{11}$ state after post-selection, which occurs with absolute probability $p_{11} = 2 p_{\rm dc} (1-p_{\rm dc})p_{\rm loss}^2 p_1^2$. As we wish to examine the regime where $p_{\rm dc}$ is comparable to or less than the probability of photon loss not occuring, it will be convenient to introduce the constant $d = \frac{p_{\rm dc}}{1 - p_{\rm loss}}$. In the regime of high loss, where $p_{\rm loss} \to 1$, the probabilities become
\begin{eqnarray}
p_{00} = 2 d (1-p_{\rm loss}) (1-p_{\rm dc}) (1-r)^2\\
p_{\psi + \phi} = 4 d (1-p_{\rm loss}) (1-p_{\rm dc})  r (1-r) \\
p_{11} = 2 d (1-p_{\rm loss}) (1-p_{\rm dc}) r^2.
\end{eqnarray}
Note that in the absence of dark counts the state after postselection will be $\rho_{\rm raw}$, which occurs with probability $p_{\rm raw} = 2 (1-p_{\rm dc})^2 (1-p_{\rm loss})p_1 \left(1-p_1+p_{\rm loss}p_1\right)$, which in the high loss regime becomes $p_{\rm raw} = 2 (1-p_{\rm dc})^2 (1-p_{\rm loss})r$. Thus the state state of the system after dark counts are included will be given by
\begin{eqnarray}
\rho_{\rm dc} &=& \frac{p_{\rm raw} \rho_{\rm raw} + p_{00}\ket{00}\bra{00} + p_{11}\ket{11}\bra{11} + \frac{p_{\psi+\phi}}{2} \ket{\psi}\bra{\psi} + \frac{p_{\psi+\phi}}{2} \ket{\phi}\bra{\phi}}{p_{\rm raw} + p_{00} + p_{11} + p_{\psi+\phi}}\\
&=& \frac{(1-p_{\rm dc}) r \rho_{\rm raw} + d (1-r)^2 \ket{00}\bra{00} + d r^2 \ket{11}\bra{11} + d r (1-r) \ket{\psi}\bra{\psi} + d r (1-r) \ket{\phi}\bra{\phi}}{d + (1-p_{\rm dc}) r}
\end{eqnarray}
As the dark count rate in many of the current generation of experiments is already very low, we can consider this expression in the case of small $p_{\rm dc}$, in which case
\begin{eqnarray}
\rho_{\rm dc} &\approx& \frac{r \rho_{\rm raw} + d (1-r)^2 \ket{00}\bra{00} + d r^2 \ket{11}\bra{11} + d r (1-r) \ket{\psi}\bra{\psi} + d r (1-r) \ket{\phi}\bra{\phi}}{d + r}\\
&=& \frac{(1-r)(d (1-r) + r \frac{p_n}{3})}{d + r} \ket{00}\bra{00} + \frac{r ((d+1) r  +  (1-r)\frac{p_n}{3})}{d + r} \ket{11}\bra{11}\\
 &&+ \frac{r (1-r)(d+1-p_n)}{d + r} \ket{\psi}\bra{\psi} + \frac{r(1-r)(d + \frac{p_n}{3})}{d + r} \ket{\phi}\bra{\phi}
\end{eqnarray}
When phase drift between the creation of this step and the application of the EPL pair distillation step is taken into account, the state of the system will be given by $\rho_{\rm dc}' = (1-p_{\rm drift})\rho_{\rm dc} + p_{\rm drift}Z\rho_{\rm dc}Z$. Hence we have 
\begin{eqnarray}
\rho_{\rm dc}' 
&\approx& \frac{(1-r)(d (1-r) + r \frac{p_n}{3})}{d + r} \ket{00}\bra{00} + \frac{r ((d+1) r  +  (1-r)\frac{p_n}{3})}{d + r} \ket{11}\bra{11}\\
 &&+ \frac{r (1-r)\left(d+1-p_n-p_{\rm drift}(1-\frac{4}{3}p_n)\right)}{d + r} \ket{\psi}\bra{\psi}\\
&&+ \frac{r(1-r)\left(d + \frac{p_n}{3}+p_{\rm drift}(1-\frac{4}{3}p_n)\right)}{d + r} \ket{\phi}\bra{\phi}.
\end{eqnarray}
Note that the application of the EPL protocol conditioned on a (1,1) outcome is not sufficient to ensure that the output state is in subspace spanned by $\ket{\psi}$ and $\ket{\phi}$, since this outcome also occurs when one pair is in the state $\ket{00}$ and the other state is $\ket{11}$. Note however that the (1,1) outcome cannot occur when both pairs are in state $\ket{00}$ or both pairs are in state $\ket{11}$. Thus, the pair produced by an application of the EPL protocol to two noisy pairs is
\begin{eqnarray}
\rho_{\rm EPL} &\approx& f(r,d,p_n) \left(\ket{00}\bra{00} + \ket{11}\bra{11}\right) + g(r,d,p_n,p_{\rm drift}) \ket{\psi}\bra{\psi} + h(r,d,p_n,p_{\rm drift}) \ket{\phi}\bra{\phi}.
\end{eqnarray}
where
\begin{eqnarray}
f(r,d,p_n) &=& \frac{p_{\rm EPL}^{-1} (1-r)(d (1-r) + r \frac{p_n}{3})r ((d+1) r  +  (1-r)\frac{p_n}{3})}{(d + r)^2} \left(\ket{00}\bra{00} + \ket{11}\bra{11}\right)\\
g(r,d,p_n,p_{\rm drift})&=& \frac{p_{\rm EPL}^{-1} r^2 (1-r)^2\left(\left(d+1-p_n-p_{\rm drift}(1-\frac{4}{3}p_n)\right)^2+\left(d + \frac{p_n}{3}+p_{\rm drift}(1-\frac{4}{3}p_n)\right)^2\right)}{2(d + r)^2} \ket{\psi}\bra{\psi}\\
h(r,d,p_n,p_{\rm drift})&=& \frac{p_{\rm EPL}^{-1} r^2(1-r)^2\left(d+1-p_n-p_{\rm drift}(1-\frac{4}{3}p_n)\right)\left(d + \frac{p_n}{3}+p_{\rm drift}(1-\frac{4}{3}p_n)\right)}{(d + r)^2} \ket{\phi}\bra{\phi}
\end{eqnarray}
and
\begin{eqnarray}
p_{\rm EPL}&=&\frac{r(1-r)}{2(d+r)^2}\Bigg(4(d (1-r) + r \frac{p_n}{3})((d+1) r + (1-r)\frac{p_n}{3})+ r (1-r) \left(2d+1-\frac{2p_n}{3}\right)^2\Bigg)
\end{eqnarray}
is the probability of obtaining the (1,1) result during the EPL protocol. The total error probability is then $\epsilon = 1-\bra{\psi}\rho_{\rm EPL}\ket{\psi} = 1-g(r,d,p_n,p_{\rm drift})$. As can be seen from Figure \ref{fig:errorprob}, dark counts begin to contribute significantly once $d$ exceeds about $0.01$.

\begin{figure}
\includegraphics[width=0.78\columnwidth]{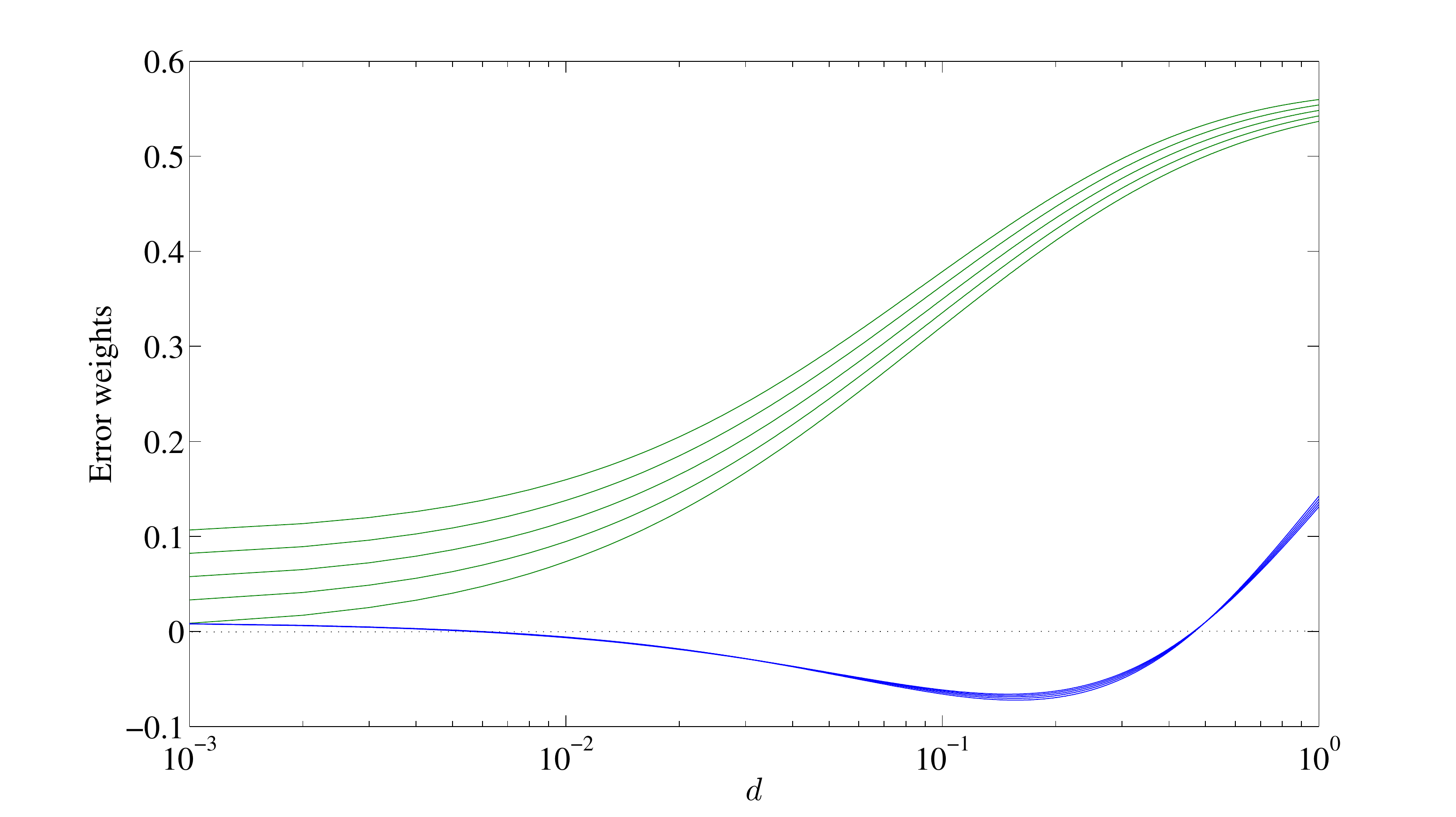}
\caption{Influence of dark counts on effective preparation and drift error probabilities for $r=\frac{1}{4}$ and $p_{\rm drift}=0.01$.The five upper lines (green) represent the effective preparation error weight ($p_n'$) corresponding from bottom to top to preparation error probabilities $p_n = \{0,0.025,0.05,0.075,0.1\}$. The lower lines (blue), indistinguishable at this scale, correspond to the effective drift error weight ($p_{\rm drift}$) for the same values of $p_n$. In the regions $d\leq0.007$ and $d\geq0.469$ both $p_n'$ and $p_{\rm drift}$ correspond to valid probabilities, and hence in these regimes dark counts are indistinguishable from other noise sources.
\label{fig:modifiederrors}}
\end{figure}

Are the errors present in this state due to dark counts are fundamentally different from those due to preparation and drift errors? Comparing $\rho_{\rm EPL}$ to the case of no dark counts, it is possible to find modified preparation and drift error weights ($p_n'$ and $p_{\rm drift}'$) such that the two states match. As $f$ is independent of $p_{\rm drift}$, the value of $p_n'$ can be obtained by solving $f(r,d,p_n) = f(r,0,p_n)$. The value for $p_{\rm drift}'$ can the be obtained by solving $g(r,d,p_n,p_{\rm drift})=g(r,d,p_n',p_{\rm drift}')$. Although $p_n'$ and $p_{\rm drift}'$ are not gauranteed to correspond to valid probabilities, for many experimentally relevant parameter ranges they do indeed take on values between zero and one, thus the effect of dark counts in these ranges is indistinguishable from drift and preparation errors. As can be seen from Figure \ref{fig:modifiederrors}, as long as the value of $d$ is kept far below $1$ (in this case around 0.01), the effective modification of the drift and preparation error rates due to dark counts is relatively small. However, as expected when $d$ approaches unity the effective error rate rapidly increases.

\begin{thebibliography}{10}

\bibitem{LucasPreP}
T.~Harty, D.T.C. Allcock, C.J. Ballance, L.~Guidoni, H.A. Janacek, N.M. Linke,
  D.N. Stacey, and D.M. Lucas.
\newblock High-fidelity preparation, gates, memory and readout of a trapped-ion
  quantum bit.
\newblock {\em arXiv:1403.1524 [quant-ph]}, 2014.

\bibitem{IonBest2Q}
C.~J. Ballance, T.~P. Harty, N.~M. Linke, and D.~M. Lucas.
\newblock High-fidelity two-qubit quantum logic gates using trapped calcium-43
  ions.
\newblock {\em arXiv:1406.5473}, 2014.

\bibitem{MartinisNew}
R.~Barends, J.~Kelly, A.~Megrant, A.~Veitia, D.~Sank, E.~Jeffrey, T.~C. White,
  J.~Mutus, A.~G. Fowler, B.~Campbell, Y.~Chen, Z.~Chen, B.~Chiaro,
  A.~Dunsworth, C.~Neill, P.~O`Malley, P.~Roushan, A.~Vainsencher, J.~Wenner,
  A.~N. Korotkov, A.~N. Cleland, and J.~M. Martinis.
\newblock Logic gates at the surface code threshold: Superconducting qubits
  poised for fault-tolerant quantum computing.
\newblock {\em arXiv:1402.4848 [quant-ph]}, 2014.

\bibitem{Wrachtrup14}
F.~Dolde, V.~Bergholm, Y.~Wang, I.~Jakobi, B.~Naydenov, S.~Pezzagna, J.~Meijer,
  F.~Jelezko, P.~Neumann, T.~Schulte-Herbr\"uggen, J.~Biamonte, and
  J.~Wrachtrup.
\newblock High-fidelity spin entanglement using optimal control.
\newblock {\em Nature Communications}, 5:3371, 2014.

\bibitem{distantNuclei}
T.~H. Taminiau, J.~Cramer, T.~van~der Sar, V.~V. Dobrovitski, and R.~Hanson.
\newblock Universal control and error correction in multi-qubit spin registers
  in diamond.
\newblock {\em Nat Nano}, 9:171--176, 2014.

\bibitem{monogEnt}
Dong Yang.
\newblock A simple proof of monogamy of entanglement.
\newblock {\em Physics Letters A}, 360:249, 2006.

\bibitem{MUSIQC}
C.~Monroe, R.~Raussendorf, A.~Ruthven, K.R. Brown, P.~Maunz, L.-M. Duan, and
  J.~Kim.
\newblock Large scale modular quantum computer architecture with atomic memory
  and photonic interconnects.
\newblock {\em Phys. Rev. A}, 89:022317, 2014.

\bibitem{monroeNew}
D.~Hucul, I.V. Inlek, G.~Vittorini, C.~Crocker, S.~Debnath, S.M. Clark, and C.l
  Monroe.
\newblock Modular entanglement of atomic qubits using both photons and phonons.
\newblock {\em arXiv:1403.3696 [quant-ph]}, 2014.

\bibitem{SQDcavityCouple}
L.~DiCarlo, J.M. Chow, J.M. Gambetta, L.S. Bishop, B.R. Johnson, D.I. Schuster,
  J.~Majer, A.~Blais, L.~Frunzio, S.M. Girvin, and R.J. Schoelkopf.
\newblock Demonstration of two-qubit algorithms with a superconducting quantum
  processor.
\newblock {\em Nature}, 460:240, 2009.

\bibitem{Roch_RemoteEntanglement}
N.~Roch, M.E. Schwartz, F.~Motzoi, C.~Macklin, R.~Vijay, A.W. Eddins, A.N.
  Korotkov, K.B. Whaley, M.~Sarovar, and I.~Siddiqi.
\newblock Observation of measurement-induced entanglement and quantum
  trajectories of remote superconducting qubits.
\newblock {\em Phys. Rev. Lett.}, 112:170501, 2014.

\bibitem{BERNIEN_2012nvEntanglement}
H.~Bernien, B.~Hensen, W.~Pfaff, G.~Koolstra, M.~S. Blok, L.~Robledo, T.~H.
  Taminiau, M.~Markham, D.~J. Twitchen, L.~Childress, and R.~Hanson.
\newblock Heralded entanglement between solid-state qubits separated by 3
  meters.
\newblock {\em Nature}, 497:86--90, 2013.

\bibitem{DelftTeleport}
W.~{\em et al} Pfaff.
\newblock Unconditional quantum teleportation between distant solid-state
  qubits.
\newblock {\em Science DOI: 10.1126/science.1253512}, 2014.

\bibitem{BRIEGEL_entanglementPurification}
W.~D\"ur and H.-J. Briegel.
\newblock Entanglement purification for quantum computation.
\newblock {\em Phys. Rev. Lett.}, 90:067901, 2003.

\bibitem{EarlPRA}
E.~Campbell.
\newblock Distributed quantum-information processing with minimal local
  resources.
\newblock {\em Physical Review A}, 76:040302, 2007.

\bibitem{FUJII_doubleSelection}
K.~Fujii and K.~Yamamoto.
\newblock Entanglement purification with double selection.
\newblock {\em Phys. Rev. A}, 80:042308, Oct 2009.

\bibitem{FUJII_distributedArchitechture}
K.~Fujii, T.~Yamamoto, M.~Koashi, and N.~Imoto.
\newblock A distributed architecture for scalable quantum computation with
  realistically noisy devices.
\newblock {\em pre-print}, arXiv:1202.6588v1, 2012.

\bibitem{BriegelNew}
M.~Zwerger, H.~J. Briegel, and W.~D\"ur.
\newblock Hybrid architecture for encoded measurement-based quantum
  computation.
\newblock {\em arXiv:1308.4561 [quant-ph]}, 2013.

\bibitem{FOWLER_2009high}
A.~G. Fowler, A.~M. Stephens, and P.~Groszkowski.
\newblock High-threshold universal quantum computation on the surface code.
\newblock {\em Physical Review A}, 80(5):052312, 2009.

\bibitem{WANG_QCwithNNinteractionsAndErrorRatesOverOnePercent}
D.~S. Wang, A.~G. Fowler, and L.~C.~L. Hollenberg.
\newblock Quantum computing with nearest neighbor interactions and error rates
  over 1\%.
\newblock {\em Phys. Rev.}, 83(2), 2011.

\bibitem{nickerson_topological}
N.~H. Nickerson, Y.~Li, and S.~C. Benjamin.
\newblock Topological quantum computing with a very noisy network and error
  rates approaching one percent.
\newblock {\em Nat. Commun.}, 2013.

\bibitem{Kitaev97}
A.~Kitaev.
\newblock Quantum error correction with imperfect gates.
\newblock In {\em Proceeding of the Third International Conference on Quantum
  Communication and Measurement}, 1997.

\bibitem{Kitaev03fault-tolerantquantum}
A.~Y. Kitaev.
\newblock Fault-tolerant quantum computation by anyons.
\newblock {\em Annals Phys.}, 303:2--30, 2003.

\bibitem{DENNIS_topolQuantumMem}
E.~Dennis, A.~Kitaev, A.~Landahl, and J.~Preskill.
\newblock Topological quantum memory.
\newblock {\em J. Math. Phys.}, 43(9):4452, 2002.

\bibitem{HORSMAN_Lattice_surgery}
C.~Horsman, A.~G. Fowler, S.~Devitt, and R.~Van~Meter.
\newblock Surface code quantum computing by lattice surgery.
\newblock {\em New Journal of Physics}, 14(12):123011, 2012.

\bibitem{BRAVYI_magicStates}
S.~Bravyi and A.~Kitaev.
\newblock Universal quantum computation with ideal clifford gates and noisy
  ancillas.
\newblock {\em Phys. Rev. A}, 71:022316, Feb 2005.

\bibitem{BARRETT_KOK}
S.~D. Barrett and P.~Kok.
\newblock Efficient high-fidelity quantum computation using matter qubits and
  linear optics.
\newblock {\em Phys. Rev. A}, 71:060310, Jun 2005.

\bibitem{ReisererNature2014}
A.~Reiserer, N.~Kalb, G.~Rempe, and S.~Ritter.
\newblock A quantum gate between a flying optical photon and a single trapped
  atom.
\newblock {\em Nature}, 508:237, 2014.

\bibitem{steinerArxiv}
M.~M.~Steiner, H.~M. Meyer, J.~Reichel, and M.~K\"ohl.
\newblock Photon emission and absorption of a single ion coupled to an optical
  fiber-cavity.
\newblock {\em arXiv:1407.6036}, 2014.

\bibitem{CAMPBELL_extremePhotonLoss}
E.~T. Campbell and S.~C. Benjamin.
\newblock Measurement-based entanglement under conditions of extreme photon
  loss.
\newblock {\em Phys. Rev. Lett.}, 101:130502, Sep 2008.

\bibitem{brokerClient}
S.~C. Benjamin, D.~E. Browne, J.~Fitzsimons, and J.~J.~L Morton.
\newblock Brokered graph-state quantum computation.
\newblock {\em New Journal of Physics}, 8(8):141, 2006.

\bibitem{Bravyi_RareEvents}
S.~Bravyi and A.~Vargo.
\newblock Simulation of rare events in quantum error correction.
\newblock {\em Phys. Rev. A}, 88:062308, Dec 2013.

\bibitem{Watson_overheads}
F.H.E. Watson and S.D. Barrett.
\newblock Estimating overheads for topological quantum codes.
\newblock {\em arXiv:1402.4848 [quant-ph]}, 2014.

\bibitem{Jamiolkowskii}
A.~Jamiolkowski.
\newblock Linear transformations which preserve trace and positive
  semidefiniteness of operators.
\newblock {\em Reports on Mathematical Physics}, 3(4):275 -- 278, 1972.

\bibitem{edmonds1965paths}
Jack Edmonds.
\newblock Paths, trees, and flowers.
\newblock {\em Canadian Journal of mathematics}, 17(3):449--467, 1965.

\bibitem{kolmogorov2009blossom}
V.~Kolmogorov.
\newblock Blossom {V}: a new implementation of a minimum cost perfect matching
  algorithm.
\newblock {\em Math. Prog. Comp.}, 1:43, 2009.

\bibitem{WANG_planarcodesl}
D.~S. Wang, A.~G. Fowler, A.~M. Stephens, and L.~C.~L. Hollenberg.
\newblock Threshold error rates for the toric and planar codes.
\newblock {\em Quantum Information and Computation}, 10:456--469, 2010.

\end{thebibliography}
\end{document}